\documentclass[12pt,a4paper]{article}
\usepackage[T1]{fontenc} 
\usepackage{physics}
\usepackage{float}
\usepackage{lmodern}
\usepackage[numbers,sort&compress]{natbib}
\usepackage{epsfig}
\usepackage{multirow}
\usepackage{inputenc}
\usepackage{hyperref}
\usepackage{graphicx}
\usepackage{wrapfig}
\usepackage{amsmath}
\usepackage{xparse}
\usepackage{amsfonts}
\usepackage{amssymb}
\usepackage{amsmath}
\usepackage{relsize} 
\usepackage{physics}
\usepackage{subcaption}
\usepackage{xcolor}
\usepackage{quotes}
\usepackage{calligra}
\usepackage{mathrsfs}
\DeclareRobustCommand{\rc}{\text{\Large\calligra r}}
\NewDocumentCommand{\tens}{t_}
{%
	\IfBooleanTF{#1}
	{\tensop}
	{\otimes}%
}
\NewDocumentCommand{\tensop}{m}
{%
	\mathbin{\mathop{\otimes}\displaylimits_{#1}}%
}
\usepackage{changepage} 

\usepackage{afterpage}

\usepackage{placeins}
\newcommand{\z}{\Bar{z}}

\newcommand{\paren}[1]{\left(#1\right)}
\newcommand{\m}{\Tilde{m}}
\newcommand{\C}{(l^2+\Bar{z}^2)}
\newcommand{\ci}{(l^2+z_i^2)}
\newcommand{\G}{\frac{z_i^2\sqrt{\tilde{m}}}{l^2+z_i^2}\left(z_i^2+\frac{z_iz_t}{2}\right)}
\newcommand{\T}[1]{\tilde{#1}}
\newcommand{\ec}{\mathscr{K}}

\usepackage{wrapfig}
\usepackage[nottoc,notlof,notlot]{tocbibind} 
\usepackage{dblfloatfix}
\usepackage{authblk}
\usepackage[top=1in, bottom=2cm, left=2cm, right=2cm]{geometry}
\usepackage{mathrsfs}

\makeatletter
\@tfor\@tempa:=ABCDEFGHIJKLMNOPQRSTUVWXYZ\do{%
  \expandafter\edef\csname scr\@tempa\endcsname{\noexpand\mathscr{\@tempa}}%
}

\title{Entanglement Entropy and Complexity of Multicomponent Universe from Holography}
\author{\bf  \normalsize{Ritam Mahanta} \thanks{ritam.mahanta@bose.res.in}}
\author{\bf  Gopinath Guin \thanks{gopinath.guin@bose.res.in}}
\author{\bf  \normalsize{Souvik Paul} \thanks{souvik.paul@bose.res.in}}
\author{\bf Sunandan Gangopadhyay \thanks{sunandan.gangopadhyay@bose.res.in}}
\affil{\large{Department of Astrophysics and High Energy Physics,\\
	S.N.~Bose National Centre for Basic Sciences,\\
	Salt Lake, Kolkata 700106, India}}

\date{}
\begin{document}
\maketitle
\begin{abstract}
\footnotesize \noindent Recent studies in \cite{Park:2020jio,Paul:2025gpk} have calculated various holographic information-theoretic quantities of the four-dimensional FLRW universe for different matter-dominated eras using the braneworld model of cosmology. These studies are done for a single matter component, which is a good toy model for understanding the entanglement properties of the universe. However, for a more realistic model, one should consider a scenario where our universe has coexisting matter components like radiation-dark matter or radiation-exotic matter, etc. In this work, we have presented a systematic way to study various holographic information-theoretic quantities, namely, entanglement entropy and complexity, of the FLRW universe in the presence of coexisting matter components. We have shown that the black brane geometry in the presence of $p$-brane gas indeed supports the existence of a universe with two-component matter sources. The second Israel junction condition, along with the Ryu-Takayanagi formula, is used to compute the time-dependent holographic entanglement entropy of the universe with coexisting radiation-dark matter and radiation-exotic matter. The expression of the time-dependent volume complexity is also evaluated in these scenarios. For both universes, these information-theoretic quantities show a clear radiation dependence in the early time and matter and exotic matter dominance in the late time, which is consistent with the thermal history of the universe \cite{WMAP:2010qai,WMAP:2010sfg,Planck:2014loa,Planck:2018vyg}.
\end{abstract}
\newpage
\tableofcontents
\section{Introduction}
	Entanglement entropy (EE) is an important information-theoretic quantity and has been an active area of research, gaining attention to study various physical systems and their various quantum features. It turns out that entanglement entropy is a very good measure of the entanglement of pure states. The entanglement entropy in quantum information theory can be realised in the following way by defining a state $\ket\psi$ for the whole system and then dividing it into two subsystems (say $\scrA$ and $\scrB$), one can calculate the entanglement entropy of any of the subsystems given by von Neumann\cite{von2013mathematische}. The von Neumann entropy for the reduced density matrix of that subsystem $\scrA$ is given by
\begin{equation}
    S_{\mathscr{A}}=-Tr\paren{\rho_\scrA \ln{\rho_\scrA}}~.
\end{equation}
Here, $\rho_\scrA$ is the reduced density matrix of the subsystem $\scrA$, which is calculated by tracing over the degrees of freedom of the subsystem $\scrB$. Mathematically, it is given by
\begin{equation}
    \rho_\scrA=Tr_\scrB\paren{\ket{\psi}\bra{\psi}}~.
\end{equation}
Even knowing the importance of quantum entanglement, it is very hard to compute entanglement entropy except for the simplest systems. The prescription to calculate the entanglement entropy in quantum field theory is called the replica trick. In \cite{Calabrese:2004eu}, the exact calculation of the entanglement entropy has been carried out for subsystems on different topological manifolds (for example, CFT on a finite strip and CFT on a circle) for $(1+1)$-dimensions. In higher-dimensional ($d>2$) field theory, it is very hard to compute the entanglement entropy.   Again, studying strongly coupled quantum field theory using conventional methods (perturbative methods in CFT) is extremely difficult, and for higher dimensions, it is almost impossible. In this stuck scenario, Maldacena came up with a genius idea, proposing a holographic duality between strongly coupled gauge theories with weakly coupled gravity theory\cite{Maldacena:1997re}. This theory is called AdS/CFT correspondence \cite{Maldacena:1997re,Gubser:1998bc,Witten:1998qj}. According to this conjecture, one considers a $d$-dimensional CFT theory living on the boundary of a $d+1$-dimensional gravity theory in AdS space, which is easier to work with.  This conjecture is widely used in several branches of physics, for example black hole physics \cite{Paul:2024rto, Gregory:2004vt,Tanaka:2002rb,Emparan:2002px,Bak:2006nh,Rovelli:1997na,Silva:2023ieb,Engelhardt:2015gla}, quantum information theory \cite{Ryu:2006bv,Ryu:2006ef,Nishioka:2009un,Takayanagi:2017knl,Jain:2017aqk,Ghasemi:2021jiy,Jokela:2019ebz,Caceres:2018blh,Mishra:2016yor,Karar:2019bwy,Chowdhury:2021idy,ChowdhuryRoy:2022dgo,Nguyen:2017yqw,Kusuki:2019zsp,Chaturvedi:2016rcn,Saha:2021kwq,Parihar:2025oxz}, QCD \cite{Csaki:2008dt,Andreev:2006ct,Karch:2006pv,Kruczenski:2004me,Erlich:2005qh,Panero:2009tv}, condensed matter physics \cite{Hartnoll:2008kx,Hartnoll:2008vx,Herzog:2009xv,Gangopadhyay:2012am,Li:2011xja,Horowitz:2010gk,Paul:2024lmd,Paul:2025apr}, cosmology \cite{Paul:2025gpk,Park:2020jio,McFadden:2009fg,Banks:2001px,Bak:1998vj,HERTOG2007397,Lepe:2008ka,Nastase:2019rsn,Waddell:2022fbn,Betzios:2020zaj}, etc. \\
Although entanglement entropy (that is, the von Neumann entropy) is a good measure of entanglement for pure states, due to the presence of classical correlations, it is not a good measure for mixed states. For mixed states, people calculate other entanglement measures such as entanglement purification, entanglement negativity and mutual information, etc. In the literature, other mixed state measures like mixed state entanglement measures, quantum complexity, etc., have been calculated. In holography, complexity is a measure used to characterise the growth of the Einstein-Rosen Bridge (ERB) connecting the two sides of the Penrose diagram for an eternal AdS black hole, particularly after the scrambling time when the dynamics of the quantum states appear to stop. If there is a predefined initial state $\ket{\psi_I}$ and a target final state $\ket{\psi_F}$, then the complexity is defined as the minimum number of elementary unitary operations (called gates) which are elements of a fixed universal gate set ($\scrG$) required to connect the initial state to the target final state. Mathematically \cite{nielsen2010quantum,Baiguera:2025dkc}
\begin{equation}
    C_{state}\paren{\ket{\psi_F}; \ket{\psi_I}}=min
_\scrU C_{unitary}[\scrU]=min\{L|\scrU_L\scrU_{L-1}...\scrU_1,\scrU_i\in\scrG\}~
\end{equation} 
where $\ket{\psi_F}=\scrU\ket{\psi_I}$. In holography, there are several conjectures about how to measure and compute complexity, for example, complexity proportional to the length of the ERB \cite{Susskind:2014rva,Susskind:2018pmk,Susskind:2014moa}, complexity proportional to the volume co-dimension one surface bounded by the fixed time slices (spatial slices) of the boundary CFT \cite{Susskind:2014rva,Susskind:2014moa}, etc. In this article, we have used the complexity proportional to the volume conjecture to calculate the complexity of the FLRW (Friedmann–Lemaître–Robertson–Walker \cite{Friedman:1922kd,Friedmann:1924bb,Lemaitre:1931zza,Lemaitre:1933gd,Robertson:1935jpx,Robertson:1935zz,Walker:1937qxv}) universe using holographic duality. The FLRW metric is a unique solution of the Einstein equation for a homogeneous and isotropic universe. At the beginning, Einstein thought of the universe as a static one, and to balance the attractive force of the matter content, he introduced the Cosmological constant term; however, there were serious stability issues with this model, which were pointed out by Eddington and others. Later, the discovery of the expanding universe by  Hubble led Einstein to comment that introducing the cosmological constant was his greatest blunder. However, modern cosmology accepts the cosmological constant term as a source of an accelerating universe. In an expanding universe, the gauge field behaves dynamically, and at the age of a few microseconds, the universe had a temperature comparable to the critical temperature of QCD, creating a perfect environment for strongly coupled processes to happen. \\
Calculating entanglement entropy using a holographic setup requires the Ryu-Takayanagi (RT) formula\cite{Ryu:2006bv,Ryu:2006ef}. However, for a time-dependent system, the Hubeny-Rangamani-Takayanagi (HRT) prescription \cite{Hubeny:2007xt} should be used instead of RT. In the HRT formalism, the minimal surface extends into the time direction as well. However, HRT can only be applied to a limited number of simple systems to obtain the exact form of entanglement entropy. 
Although HRT is a precise method for analysing time-dependent systems, at a fixed moment in time, the RT formalism can be employed to obtain the leading-order contribution of HRT in the ultraviolet (UV) limit \cite{Hubeny:2007xt}. Nonetheless, challenges remain, such as accounting for higher-order contributions that could influence the late-time behavior and the computation of entanglement entropy in universes characterised by AdS and dS geometries undergoing power-law expansion.\\
To calculate the time dependence of various information-theoretic quantities of our universe, the brane world (Randall-Sundrum I) model \cite{Randall:1999ee,Randall:1999vf,Chamblin:1999by,Chamblin:1999ya,Brax:2003fv,Brax:2004xh,Flanagan:1999dc,Coley:2001ab} has been used in the literature, taking the universe to be living in a four-dimensional brane, and the time evolution of the universe comes from the movement of the brane in the bulk direction. At first, the brane world technique was introduced to address the hierarchy problem of two different energy scales late, but it has been improved and has been used to explain a universe going through inflation and graceful exit \cite{Kraus:1999it,Park:2000ga,Papantonopoulos:2004au,Okuyama:2004in,Yoshiguchi:2004cb,Chang:2004xs,Bronnikov:2025ppq}. In this direction, the time dependence of the entanglement entropy for the FLRW universe was studied exactly using p-brane gas geometry in \cite{Park:2020jio}, where the author had taken a five-dimensional bulk geometry without any bulk matter field to get radiation dominated universe and the addition of string cloud in the background metric gives there the dark matter dominated universe for early and late time era. In \cite{Paul:2025gpk}, authors have studied the HEE of matter, radiation, and exotic matter-dominated universe separately using the perturbative method, where they have shown that by that procedure the form of the HEE matches asymptotically with the exact result produced in \cite{Park:2020jio} for both in early and late time. Holographic complexity has also been calculated using the same perturbative method for the early and late time limits. In all this procedure, RT formalism has been used by considering the expansion of the universe dual to the movement of the brane in the bulk direction, and HEE is not an explicit function of time there. The time dependence is entering through the brane position, and the time-dependent brane position is calculated using the second Israel junction condition. But in all this literature, a study has been done for a single-component universe, surely, which is not our universe. Some interesting studies regarding various information-theoretic quantities for the braneworld cosmological model can be found in \cite{Geng:2020fxl,Geng:2021wcq,Geng:2023iqd,Park:2021wep,Iwashita:2006zj,Kushihara:2021fbr,Feng:2023krm,Basu:2025sqk,Bhattacharya:2023drv}. Also, the effect of anisotropy in the context of correlation and holographic information theoretic measures of the universe has been studied in detail in \cite{Narayan:2024fcp,Jiang:2025ktt,Noumi:2025cup,Giantsos:2022qdd,Giataganas:2021cwg,Carrillo-Gonzalez:2020ejs,Engelhardt:2014mea,Marcori:2016oyn,Chatterjee:2016bhj,Banerjee:2015fua,doi:10.1142/S0217732325502335} \\
This motivates us to study a universe where there are multiple components and the effect of the subdominant component in the particular dominated universe. This approach is justified by observations of the cosmic microwave background radiation \cite{WMAP:2010sfg,Planck:2014loa}, which indicate that, in the early universe, radiation was the dominant component, and following the recombination period, matter took over as the primary constituent. Therefore, studying such a multi-component universe provides a more accurate depiction of cosmic evolution. So it is more accurate to study such kind of universe. The main challenge is that the arbitrary inclusion of such a component does not satisfy the Einstein equation. But the inclusion of one one-dimensional string on a brane with the five-dimensional AdS black brane generates a blackening factor that contains both matter and radiation. We further investigated a universe where there is radiation and exotic matter. To get such behaviour, here we have considered the AdS black brane with a string two-brane to get a blackening factor which contains the effect of radiation and exotic matter. \\
This article is arranged in the following way. First, we will explicitly derive the Israel junction condition for a general metric by varying the Einstein-Hilbert metric with Gibbons–Hawking–York (GHY) boundary term \cite{Gibbons:1976ue,Hawking:1995fd,York:1972sj,Deruelle:2017xel} and demanding the stress energy tensor of the brane as the difference of the stress tensor on the two sides of the brane. Then we have derived the metric for different matter components of the universe. Following this, we have moved on to calculate the entanglement entropy for the pure AdS manifold for early and late times separately using the perturbative technique employed in \cite{Paul:2025gpk}. Using the same technique, we then studied the HEE for a realistic universe having both matter and radiation in early and late times. We have also carried out a detailed analysis of the study of time-dependent HEE in a universe with coexisting radiation and matter in the intermediate regime. The next subsection is dedicated to calculating the same for a universe with radiation and exotic matter. In section \eqref{complexity section}, we have focused on the calculation of complexity, starting with pure AdS, followed by radiation-matter universe and radiation-exotic matter universe in both early and late times. The volume complexity for a universe with coexisting radiation-matter is also computed in the intermediate regime. Before concluding, a discussion has been given about the results and their implications. 
\section{Braneworld model and Israel junction condition}
The braneworld model of cosmology (in our case, the RS-II braneworld model) states that our four-dimensional FLRW universe is situated on a brane. This brane is embedded in one higher-dimensional spacetime, that is, a five-dimensional spacetime. According to this model, the expansion of the universe is analogous to the radial motion of the brane along the radial bulk direction. In this model, it is assumed that the brane sets the boundary between two different bulk spacetimes (say $\mathcal{M}_+$ and $\mathcal{M}_-$). Although for simplicity, we will take the same form of the spacetime metric on both sides of the brane. We will use this braneworld model along with the Ryu-Takayanagi formula \cite{Ryu:2006bv,Ryu:2006ef} to compute the entanglement entropy and complexity of some part of our universe. To incorporate the time-dependence in our results, we will use something called the Israel junction condition. In the literature, there are two Israel junction conditions. The first one says that the induced metric on both sides of the brane must match on the brane surface to obtain a unique metric on the brane\cite{Israel:1966rt}. \\
We will now briefly derive the second Israel junction condition. Let us start with the Einstein-Hilbert action along with the GHY boundary term
\begin{equation}
    S=\frac{1}{16\pi G_5}\int d^5x \sqrt{-g} (\mathcal{R}-2\Lambda)+\frac{1}{8\pi G_4}\int d^4x \sqrt{-h}\ec = S_{EH}+S_{GHY}~.
\end{equation}
In the above action $\mathcal{R}$ is the Ricci scalar, $\Lambda$ is the cosmological constant, $\ec$ is the trace of the extrinsic curvature tensor, $g$ and $h$ are the determinants of the total spacetime metric and the boundary metric, respectively.
With this action in hand, we will proceed to derive an expression for the bulk canonical momenta. In order to do so, we will first vary the Einstein-Hilbert action with respect to the total spacetime metric $g_{\mu\nu}$ gives 
\begin{equation}
    \delta S_{EH}=\frac{1}{16\pi G_{5}}\int d^5 x \Bigg[\sqrt{-g}(R_{\mu\nu}-\frac{1}{2}Rg_{\mu\nu}-\Lambda g_{\mu\nu})\delta g^{\mu\nu}+\sqrt{-g}\delta R_{\mu\nu}g^{\mu\nu}\Bigg]~.
\end{equation}
The first term in the above equation gives the well-known Einstein's equation. Now, let us focus on the second term
\begin{equation}
    \frac{1}{16\pi G_{5}}\int_{\mathcal{M}}d^5x \sqrt{-g}\delta R_{\mu\nu}g^{\mu\nu}~.
\end{equation}
In order to vary the Ricci tensor, we can use the Platini identity, which is given by
\begin{align}\label{var Rmunu}
    &\delta R_{\mu\nu}=\nabla_{\lambda}\delta \Gamma^{\lambda}_{\mu\nu}-\nabla_{\nu}\delta \Gamma^{\lambda}_{\lambda\mu}\nonumber\\
    &\implies g^{\mu\nu}\delta R_{\mu\nu}=\nabla_{\lambda}(g^{\mu\nu}\delta \Gamma^{\lambda}_{\mu\nu})-\nabla_{\nu}(g^{\mu\nu}\delta \Gamma^{\lambda}_{\lambda\mu})~.
\end{align}
In the last line, we have used the metric compatibility property. In the above expression, the metric variation of the Christoffel symbol is given by the following formula
\begin{equation}
    \delta\Gamma^{\lambda}_{\mu\nu}=\frac{1}{2}g^{\lambda\rho}\left(\nabla_{\mu}\delta g_{\nu\rho}+\nabla_{\nu}\delta g_{\mu\rho}-\nabla_{\rho}\delta g_{\mu\nu}\right)~.
\end{equation}
After substituting this formula for the variation of the Christoffel symbol in eq.\eqref{var Rmunu} and after doing a little bit of algebra, we obtain
\begin{equation}
    g^{\mu\nu}\delta R_{\mu\nu} = \nabla_{\lambda}(g^{\mu\nu}\delta \Gamma^{\lambda}_{\mu\nu}-g^{\mu\lambda}\delta \Gamma^{\rho}_{\mu\rho})~.
\end{equation}
The above term is a total derivative term in the bulk action; therefore, we can use Gauss's divergence theorem, which reads
\begin{equation}
    \frac{1}{16 \pi G_{5}}\int_{\mathcal{M}}d^5x \sqrt{-g}\nabla_{\lambda}(g^{\mu\nu}\delta \Gamma^{\lambda}_{\mu\nu}-g^{\mu\lambda}\delta \Gamma^{\rho}_{\mu\rho})=\frac{1}{16\pi G_{4}}\int_{\partial\mathcal{M}}d^4x\sqrt{-h}n_{\lambda}(g^{\mu\nu}\delta \Gamma^{\lambda}_{\mu\nu}-g^{\mu\lambda}\delta \Gamma^{\rho}_{\mu\rho})
\end{equation}
where $G_5$ is the Newton constant, in five dimensions and related to the four-dimensional counterpart $G_4$ as $G_5=2\pi a G_4$, $a$ being the radius of the compactified extra dimension on a circle \cite{Becker:2006dvp}. As $a$ is a constant, we have defined $2\pi a=1$. $h$ is the determinant of the boundary induced metric, $n_{\lambda}$ is some unit normal vector on the boundary and $\partial\mathcal{M}$ is the boundary manifold. Upon further simplification, one can show that
\begin{align}\label{eq bdy var EH term}
    &\frac{1}{16\pi G_{4}}\int_{\partial\mathcal{M}}d^4x\sqrt{-h}n_{\lambda}(g^{\mu\nu}\delta \Gamma^{\lambda}_{\mu\nu}-g^{\mu\lambda}\delta \Gamma^{\rho}_{\mu\rho})\nonumber\\&=\frac{1}{16\pi G_{4}}\int_{\partial\mathcal{M}}d^4x\sqrt{-h}n^{P}g^{MN}(\nabla_{M}\delta g_{NP}-\nabla_{P}\delta g_{MN})~.
\end{align}
Now the boundary metric can be written in terms of the total spacetime metric as \\ 
\begin{equation}
    h^{MN}=g^{MN}-n^M n^N
\end{equation}
Using this relation, one can rewrite eq.\eqref{eq bdy var EH term} as follows
\begin{align}
    &\frac{1}{16\pi G_{4}}\int_{\partial\mathcal{M}}d^4x\sqrt{-h}n^{P}g^{MN}(\nabla_{M}\delta g_{NP}-\nabla_{P}\delta g_{MN})\nonumber\\
    &=\frac{1}{16\pi G_{4}}\int_{\partial\mathcal{M}}d^4x\sqrt{-h}n^{P}h^{MN}(\nabla_{M}\delta g_{NP}-\nabla_{P}\delta g_{MN})\nonumber\\&+\frac{1}{16\pi G_{4}}\int_{\partial\mathcal{M}}d^4x\sqrt{-h}n^{P}n^M n^N(\nabla_{M}\delta g_{NP}-\nabla_{P}\delta g_{MN})~.
\end{align}
The term $n^Pn^Mn^N$ in the last term of the above equation is symmetric under the swapping of index $P\leftrightarrow M$, although $(\nabla_{M}\delta g_{NP}-\nabla_{P}\delta g_{MN})$ is antisymmetric under this swapping, therefore the second term as a whole is an antisymmetric term and it has a value equals to zero. So the boundary term from the Einstein-Hilbert action becomes 
\begin{equation}
    \frac{1}{16\pi G_{4}}\int_{\partial\mathcal{M}}d^4x\sqrt{-h}n^{P}h^{MN}(\nabla_{M}\delta g_{NP}-\nabla_{P}\delta g_{MN})
\end{equation}
Now, we will further proceed to vary the GHY boundary term. The variation of the GHY term with respect to the boundary metric is given by
\begin{equation}\label{delta SGHY}
    \delta S_{GHY}=\frac{1}{8\pi G_{4}}\int_{\partial\mathcal{M}}d^4x\Bigg[-\frac{1}{2}\sqrt{-h}h_{\mu\nu}\delta h^{\mu\nu}\ec+\sqrt{-h}\delta \ec\Bigg]
\end{equation}
where $\ec$ is the trace of the extrinsic curvature tensor $\ec_{\mu\nu}=h^{M}_{\mu} h^{N}_{\nu}\nabla_{M}n_{N}$. Now we will focus on the evaluation of the last term. The variation of the trace of the extrinsic curvature is given by
\begin{equation}
    \delta \ec=-\ec^{\alpha\beta}\delta g_{\alpha\beta}-h^{PQ}n^{\sigma}\Big[\nabla_{p}\delta g_{\sigma Q}-\frac{1}{2}\nabla_{\sigma}\delta g_{PQ}\Big]+\frac{1}{2}\ec n^M n^N \delta g_{MN}~.
\end{equation}
Now using this expression of $\delta \ec$ in eq.\eqref{delta SGHY}, we obtain
\begin{align}
    \delta S_{GHY}=&\frac{1}{8\pi G_{4}}\int_{\partial\mathcal{M}}d^4x\Bigg[-\frac{\sqrt{-h}}{2}h_{\mu\nu}\delta h^{\mu\nu}\ec +\sqrt{-h}\Big\{-\ec^{\alpha\beta}\delta g_{\alpha\beta}\nonumber\\&-h^{PQ}n^{\sigma}\Big(\nabla_{p}\delta g_{\sigma Q}-\frac{1}{2}\nabla_{\sigma}\delta g_{PQ}\Big)+\frac{1}{2}\ec n^M n^N \delta g_{MN}\Big\}\Bigg]~.
\end{align}
Therefore, if we only focus on the boundary terms, the variation of the total spacetime action becomes
\begin{align}\label{EH GHY var total}
    \delta S_{EH}+\delta S_{GHY}&=\frac{1}{8\pi G_4}\int_{\partial\mathcal{M}}\sqrt{-h}\Bigg[\frac{1}{2}\ec n^M n^N \delta g_{MN}+\frac{1}{2}\ec h^{MN} \delta g_{MN}\nonumber\\&-\frac{1}{2} h^{MN} n^P \nabla_{M}\delta g_{NP}-\ec^{MN}\delta g_{MN}\Bigg]~.
\end{align}
We also know that for a vector field $\mathscr{X}^{M}$ tangential to the boundary manifold $\partial\mathcal{M}$
\begin{equation}
    \nabla_{M}\mathscr{X}^{M}=h^{MN} \nabla_{M}\mathscr{X}_{N}+n^{M}n^{N} \nabla_{M}\mathscr{X}_{N}= \T{\nabla}_{M}\mathscr{X}^{M}-n^M \mathscr{X}^{\lambda} \nabla_{M}n_{\lambda}
\end{equation}
where $\T{\nabla}$ is the covariant derivative associated with the induced metric on $\partial \mathcal{M}$.
Now let us focus on the term $h^{MN} n^P \nabla_{M}\delta g_{NP}$ in eq.\eqref{EH GHY var total}. This term can be written in the following manner
\begin{align}
    h^{MN} n^P \nabla_{M}\delta g_{NP}&=\nabla_{M}(h^{MN}n^P \delta g_{NP})-\delta g_{NP}\nabla_{M}(h^{MN}n^P)\nonumber\\
    &=\T\nabla_{M}(h^{MN}n^P \delta g_{NP})+\ec n^M n^N \delta g_{MN}-\ec^{MN}\delta g_{MN}~.
\end{align}
Using the above relation in eq.\eqref{EH GHY var total}, we finally obtain
\begin{align}
     \delta S_{EH}+\delta S_{GHY}=\frac{1}{16\pi G_4}\int_{\partial\mathcal{M}}d^4x\sqrt{-h}\Bigg[&-\frac{1}{2}\T\nabla_{M}(h^{MN}n^P \delta g_{NP})+\frac{1}{2}\ec h^{MN} \delta g_{MN}\nonumber\\&-\frac{1}{2}\ec^{MN}\delta g_{MN}\Bigg]
\end{align}
In the above expression, the total derivative term is integrated out to give
\begin{align}
    \delta S_{EH}+\delta S_{GHY}&=-\frac{1}{16\pi G_4}\int_{\partial\mathcal{M}}d^4x\sqrt{-h}\Big[\ec^{MN}-\ec h^{MN} \Big]\delta g_{MN}\nonumber\\&=-\frac{1}{16\pi G_4}\int_{\partial\mathcal{M}}d^4x\sqrt{-h}\Big[\ec^{MN}-\ec h^{MN} \Big]\delta h_{MN}~.
\end{align}
Therefore, for the bulk manifolds $\mathcal{M_+}$, and $\mathcal{M_-}$ the canonical momentum is given by
\begin{equation}
     \Pi^{(\pm)}_{MN}=-\frac{1}{16\pi G_4}\Big(\ec^{(\pm)}_{MN}-\ec^{(\pm)} h_{MN}\Big)~.
\end{equation}
\begin{figure}
    \centering
    \includegraphics[width=0.5\linewidth]{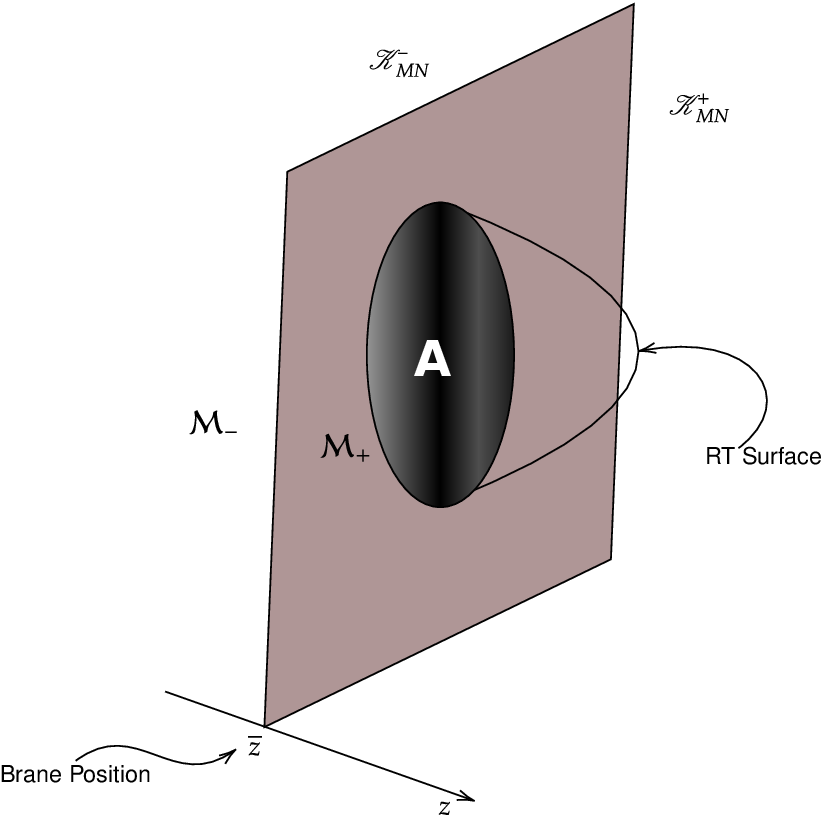}
    \caption{A schematic diagram of the braneworld model along with a circular susystem $A$ and its corresponding RT surface in the bulk.}
    \label{fig:placeholder}
\end{figure}
Now, due to the presence of $\mathcal{Z}_2$ symmetry on the brane, we can write $\ec^{(+)}_{MN}=-\ec^{(-)}_{MN}$. Therefore, denoting $\ec^{(+)}_{MN}=-\ec^{(-)}_{MN}=\ec_{MN}$, the above expression for the bulk canonical momentum can be further simplified to the following form
\begin{equation}\label{pi pm}
    \Pi_{MN}^{(\pm)}=\pm\frac{1}{16\pi G_4}\Big(\ec_{MN}-\ec h_{MN}\Big)~.
\end{equation}
Usually, it is expected that the canonical momenta on both sides of the brane must match, although due to the presence of the brane's non-vanishing energy-momentum tensor, they do not match on the brane surface. Thus, the difference of $\Pi^{(+)}$ and $\Pi^{(-)}$ must cancel the energy momentum tensor of the brane. This condition gives
\begin{equation}\label{pi plus minus brane}
    \Pi_{MN}^{(+)}-\Pi_{MN}^{(-)}=\mathfrak{T}^{(brane)}_{MN}
\end{equation}
where $\mathfrak{T}^{(brane)}_{MN}$ is the energy momentum tensor of the brane.\\
The brane's action is generally given by
\begin{equation}
    S^{(brane)}=-\frac{\mathscr{T}}{4\pi G_4}\int_{\partial\mathcal{M}}d^4x\sqrt{-h}V(\Phi)
\end{equation}
where $V$ is some potential depending on the value of the dilaton on the brane and $\frac{\mathscr{T}}{4\pi G_4}$ is the brane tension. In the ground state configuration of the brane, this action takes the following form \cite{Chamblin:1999ya}
\begin{equation}
    S^{(brane)}_{ground}=-\frac{\mathscr{T}}{4\pi G_4}\int_{\partial\mathcal{M}}d^4x\sqrt{-h}~.
\end{equation}
It is well known that the brane's energy-momentum tensor is given by the relation \cite{Chamblin:1999ya}
\begin{equation}
    \mathfrak{T}^{(brane)}_{MN}=\frac{1}{\sqrt{-h}}\frac{\delta S^{(brane)}}{\delta h_{MN}}=\frac{\mathscr{T}}{8\pi G_4}h_{MN}~.
\end{equation}
Therefore, using the above relation along with eq.\eqref{pi pm} in eq.\eqref{pi plus minus brane}, one obtains
\begin{equation}\label{jn condition with K}
    \ec_{MN}-\ec h_{MN}=\mathscr{T}h_{MN}~.
\end{equation}
We can further simplify the above relation by putting the value of $\ec$. In order to compute $\ec$, we will take the trace on both sides of eq.\eqref{jn condition with K} with respect to the boundary metric. For a four-dimensional boundary that gives 
\begin{equation}
    \ec=-\frac{4\mathscr{T}}{3}~.
\end{equation}
Now putting this value of $\ec$ in eq.\eqref{jn condition with K} leads to the following simplified equation
\begin{equation}\label{jn condition simple}
    \ec_{MN}=-\frac{\mathscr{T}}{3}h_{MN}~.
\end{equation}
We will see in a moment that this relation will be useful in computing the time-dependent radial position of the brane. \\
To calculate the extrinsic curvature tensor on the brane, one needs to first know the normal vector on the brane, which is moving in the radial direction. Let us take the five-dimensional bulk spacetime that is governed by the simple metric with no cross terms
\begin{equation}\label{metric gen}
    ds^2=-F(r)dt^2+G(r)dr^2+H(r)\delta_{ij}dx^i dx^j
\end{equation}
where $i,j=1,2,3$. We will use this spacetime metric to compute the normal vector and the extrinsic curvature on the brane. The unit normal vector for the
spacetime metric in eq.\eqref{metric gen} is given by
\begin{equation}
    n_{M}=\frac{\sqrt{FG}}{\sqrt{F-G\dot{r}^2}}\Big(\dot{r},-1,0,0,0\Big)
\end{equation}
where $\dot{r}$ is the derivative of the radial brane position with respect to the coordinate time.
We will now focus on the spatial component of the extrinsic curvature $\ec_{MN}$
\begin{equation}
    \ec_{ij}=h^{\mu}_{i} h^{\nu}_{j}\nabla_{\mu}n_{\nu}=h^{i}_{i} h^{j}_{j}\nabla_{i}n_{j}=-\Gamma^{t}_{ij}n_{t}-\Gamma^{r}_{ij}n_{r}
\end{equation}
where $\Gamma^{t}_{ij}$ and $\Gamma^{r}_{ij}$ are the Christoffel symbols. Explicit calculations show that $\Gamma^{t}_{ij}=0$ and $\Gamma^{r}_{ij}=-\frac{1}{2G}\delta_{ij}\frac{\partial H}{\partial r}$. Therefore, the spatial component of the extrinsic curvature becomes
\begin{equation}
    \ec_{ij}=-\frac{1}{2G(r)}\delta_{ij}\frac{\partial H(r)}{\partial r}\frac{\sqrt{F(r)G(r)}}{\sqrt{F(r)-G(r)\dot{r}^2}}=-h_{ij}\frac{1}{2G(r)}\frac{H^{\prime}(r)}{H(r)}\frac{\sqrt{F(r)G(r)}}{\sqrt{F(r)-G(r)\dot{r}^2}}~.
\end{equation}
where $\prime$ denotes derivative with respect to the radial coordinate $r$. Now comparing the above relation with eq.\eqref{jn condition simple}, we obtain
\begin{equation}
    \frac{H^{\prime}(r)}{H(r)}=\frac{\mathscr{T}}{3}\sqrt{\frac{G(r)}{F(r)}}\sqrt{F(r)-G(r)\dot{r}^2}~.
\end{equation}
The above expression can be rearranged to obtain the time derivative of the brane's radial position, which reads
\begin{equation}\label{dr dt}
    \dot{r}^2 =\frac{F}{G}-\frac{F}{G^2}\frac{9}{\scrT^2}\frac{H^{\prime}(r)^2}{H(r)^2}~.
\end{equation}
In order to do the cosmological analysis, we need to use the cosmological time (say $\tau$) instead of the coordinate time ($t$). In order to do the same, we will do the following transformation
\begin{equation}\label{cosmic time param}
    -d\tau^2=-F(r)dt^2+G(r)dr^2~.
\end{equation}
Under this transformation, the spacetime metric in eq.\eqref{metric gen} becomes 
\begin{equation}
    ds^2 =-d\tau^2+H(r)\delta_{ij}dx^i dx^j~.
\end{equation}
Therefore, if we identify the scale factor on the brane ($a(\tau)$) as $H(r)$, the induced metric on the brane becomes the metric of a four-dimensional FLRW universe. The identification of $H(r)$ as the scale factor will eventually reveal that the brane's radial position is nothing but the usual scale factor in the FLRW universe. We can use the transformation in eq.\eqref{cosmic time param} to recast eq.\eqref{dr dt} in a form containing the derivative of the brane's radial position with respect to the cosmic time $\tau$. This reads
\begin{equation}\label{diff_brane_position}
    \Big(\frac{dr}{d\tau}\Big)^2=\frac{\scrT^2}{9}\frac{H(r)^2}{H^{\prime}(r)^2}-\frac{1}{G(r)}~.
\end{equation}
In the upcoming sections, we will see that the above equation will be useful in determining the time-dependent radial position of the brane for various matter-dominated universes or universes with coexisting matter components.
\section{Adding matter in the universe}
Till now, we have derived an expression that governs the time-dependent radial position of the brane. However, the time-dependent radial position depends on the explicit forms of $G(r)$ and $H(r)$. The mathematical forms of these functions will depend upon the time of matter which is present in the universe. Therefore, in order to have different matter sources on the brane, we will add $p$-brane gas in the bulk spacetime. It is expected that the backreaction of this $p$-brane gas will induce different kinds of matter sources in our four-dimensional FLRW universe. \\
Let us consider the Einstein-Hilbert action with $p$-brane gas in the bulk 
\begin{equation}
    S=\frac{1}{16\pi G_5}\int d^5x \sqrt{-g} (\mathcal{R}-2\Lambda)+\mathcal{T}_p \mathcal{N}_p \int d^{p+1}\zeta \sqrt{-h}\partial^{\alpha}x_{\mu}h_{\alpha\beta}\partial^{\beta}x_{\nu}g^{\mu\nu}
\end{equation}
where $\mathcal{T}_p$ and $\mathcal{N}_p$ are the tension and number of $p$-branes in the bulk spacetime. Now we want to derive the equation of motion for the above action. In order to do so, we will vary the above action with respect to the bulk metric, which reads
\begin{equation}\label{action var}
    \delta S =\frac{1}{16\pi G_5}\int d^5 x \sqrt{-g}(\mathcal{R}_{\mu\nu}-\frac{1}{2}\mathcal{R}g_{\mu\nu} + \Lambda g_{\mu\nu})\delta g^{\mu\nu} +\mathcal{T}_{p}\mathcal{N}_{p}\int d^{p+1}\zeta \sqrt{-h}\partial^{\alpha}x_{\mu}h_{\alpha\beta}\partial^{\beta}x_{\nu}\delta g^{\mu\nu}~.
\end{equation}
At this point, we would like to mention that the bulk and the brane's world volume must have the same dimension to derive Einstein's equation of motion. In this case, the bulk and the $p$-brane's world volume have dimensions $5$ and $p+1$ respectively. Therefore, to resolve this issue, we will assume that $p$-branes are uniformly distributed in spatial directions perpendicular to the brane’s worldvolume. Let us assume that the coordinates perpendicular to the brane's world volume are $\chi^{\gamma}$, where $\chi^{\gamma}$ are the coordinates of a $(4-p)$-dimensional space with a metric $\mathcal{S}_{ab}$. Thus, the number of $p$-branes can be expressed in terms of the number density $\T n_{p}$ as follows
\begin{equation}
    \mathcal{N}_{p}=\int d^{4-p}\chi \sqrt{\mathcal{S}}\T n_p~
\end{equation}
where $\mathcal{S}$ is the determinant of the metric $\mathcal{S}_{ab}$.\\
It is important to note that because the perpendicular volume depends on the radial position, the parameter $\T n_p$ also varies with the radial location. We can also rewrite the parameter $\T n_p$ in terms of a constant number density ($n_p$) independent of the radial position, which reads
\begin{equation}
    \T n_p = \frac{n_p}{\sqrt{\mathcal{S}}}=\frac{R^{4-p}}{r^{4-p}}n_p~.
\end{equation}
Now using the above expression in eq.\eqref{action var}, we obtain the following variation of the total bulk action along with the $p$-brane gas in the bulk
\begin{align}\label{action full var}
    \delta S &=\frac{1}{16\pi G_5}\int d^5 x \sqrt{-g}(\mathcal{R}_{\mu\nu}-\frac{1}{2}\mathcal{R}g_{\mu\nu} + \Lambda g_{\mu\nu})\delta g^{\mu\nu} \nonumber\\&+\mathcal{T}_{p}\int d^{4-p}\chi\,\T n_p \sqrt{\mathcal{S}}\int d^{p+1}\zeta \sqrt{-h}\partial^{\alpha}x_{\mu}h_{\alpha\beta}\partial^{\beta}x_{\nu}\delta g^{\mu\nu}\nonumber\\&=\frac{1}{16\pi G_5}\int d^5 x \sqrt{-g}(\mathcal{R}_{\mu\nu}-\frac{1}{2}\mathcal{R}g_{\mu\nu} + \Lambda g_{\mu\nu})\delta g^{\mu\nu} \nonumber\\&+\mathcal{T}_{p}\int d^5 x \,\T n_p\sqrt{-g}\partial^{\alpha}x_{\mu}h_{\alpha\beta}\partial^{\beta}x_{\nu}\delta g^{\mu\nu}~.
\end{align}
In the above equation, we have used the fact that $\sqrt{\mathcal{S}}\sqrt{-h}=\sqrt{-g}$. \\
In five-dimensional bulk spacetime, the energy-momentum tensor corresponding to the $p$-brane gas is given by \cite{Park:2021wep}
\begin{equation}\label{EM tensor}
    T_{\mu\nu}=-\frac{\mathcal{T}_p n_p R^{4-p}}{r^{4-p}}\Big\{g_{tt},\frac{p-1}{3}g_{11},\frac{p-1}{3}g_{22},\frac{p-1}{3}g_{33},g_{uu}\Big\}~
\end{equation}
where $T_{\mu\nu}=0$ for $\mu \neq \nu$. In the above expression, $\frac{(p-1)}{3}$ denotes the average number of $p$-branes extending to one of the spatial directions. Using the above-mentioned energy-momentum tensor along with eq.\eqref{action full var}, one can obtain Einstein's equation as follows
\begin{equation}\label{Einstein eq tensor}
    \mathcal{R}_{\mu\nu}-\frac{1}{2}\mathcal{R}g_{\mu\nu} + \Lambda g_{\mu\nu}= 16\pi G_5 T_{\mu\nu}~.
\end{equation}
Now, as we are in an AdS space, due to the presence of a negative cosmological constant, the most general spacetime geometry
will be an AdS black brane. Thus, in order to solve Einstein’s equation, let us take the ansatz for the spacetime metric to be the AdS black brane, which reads
\begin{equation}
    ds^2 = \frac{r^2}{R^2}\Big(-f(r)dt^2 +\delta_{ij}dx^idx^j\Big)+\frac{R^2}{r^2f(r)}dr^2
\end{equation}
where $i,j =1,2,3$ and $f(r)$ is the lapse function of the AdS black brane geometry. Using this ansatz of the black brane metric in eq.\eqref{Einstein eq tensor}, we obtain the following set of differential equations
\begin{align}
    &\frac{r^2f(r)}{2R^4}\Big[12-12 f(r) -3rf^{'}(r)\Big]=16\pi G_{5}T_{tt}\label{Einstein eq tt}\\
    & \frac{6}{r^2}+\frac{6}{r^2f(r)}+\frac{3f^{'}(r)}{2rf(r)}=16\pi G_{5}T_{rr}~.
\end{align}
Now we will solve the differential equation in eq.\eqref{Einstein eq tt} in order to obtain the analytical form for the lapse function $f(r)$. Therefore, using the expression for $T_{tt}$ from eq.\eqref{EM tensor}, we get
\begin{equation}
    12 -\left(12f(r)+3rf^{\prime}(r)\right)=\frac{32\pi G_{5}\mathcal{T}_{p}n_p R^{6-p}}{r^{4-p}}~.
\end{equation}
Solving the above first-order differential equation, we finally obtain
\begin{equation}\label{lapse fn total raw}
    f(r)=1-\frac{32\pi G_5}{3}\mathcal{T}_{p}n_p \frac{R^{6-p}}{r^{4-p}}-\frac{C}{r^4}=1- \frac{\rho_p}{r^{4-p}}-\frac{C}{r^4}
\end{equation}
where $C$ is some arbitrary integration constant and $\rho_p$ is the energy density of the string gas given by
\begin{equation}
    \rho_p=\frac{32 \pi G_5}{3}\mathcal{T}_pn_pR^{6-p}~.
\end{equation}
We now make some important observations. The last term $\frac{C}{r^4}$ is like a Schwarzschild term, and the term $\frac{\rho_p}{r^{4-p}}$ is completely due to the backreaction of the string gas in the bulk spacetime. We can easily identify $C$ as the ADM mass \cite{Arnowitt:1959ah} of the black hole, and from now on, we will denote it by $m$. Therefore, the lapse function can be rewritten as 
\begin{equation}\label{lapse fn total}
    f(r)=1- \frac{\rho_p}{r^{4-p}}-\frac{m}{r^4}~.
\end{equation}
\\
In previous studies \cite{Park:2020jio,Paul:2025gpk}, it was shown that in the absence of any string gas, the lapse function of the black brane becomes the lapse function of a Schwarzschild black hole. The back reaction of this bulk object is responsible for the presence of radiation in the four-dimensional FLRW universe on the brane. In several previous studies in the context of brane world cosmology, the effects of radiation and matter on the brane have been analyzed by solving the Einstein field equations of general relativity \cite{Mukohyama:1999qx, Shiromizu:1999wj,Binetruy:1999ut,Savonije:2001nd}. These studies have also shown that in the absence of any Schwarzschild-like term in the lapse function, different kinds of $p$-brane gas configurations give rise to dark matter and exotic matter. Although if someone does not set the Schwarzschild-like term to be zero, one can study the effect of a Schwarzschild black hole surrounded by string gas ($p$-brane gas) on the FLRW universe. This technique provides a systematic way to study the RS-II braneworld model of cosmology in the presence of coexisting matter sources like radiation-dark matter, radiation-exotic matter, etc. We would also like to mention that the structure of the lapse function in eq.\eqref{lapse fn total} suggests that for the coexistence of matter in the FLRW universe, radiation is always present and depending upon various $p$-brane configurations, we have the other matter component in the universe.\\
Now our main task is to find the time-dependent brane position for universes with coexisting matter components. In order to do so, we will use the black brane lapse function of eq.\eqref{lapse fn total} and use it in the Israel junction condition in eq.\eqref{diff_brane_position}. In the next part of this section, we have explicitly computed the time-dependent brane position for universes with coexisting matter components, like a universe with coexisting radiation, dark matter and radiation with some exotic matter. 
\subsection*{Finding time dependence of the brane position}
Now we will explicitly compute the time-dependent radial position of the brane for universes with co-existing radiation-matter and radiation-exotic matter. We will also show the early and late time calculations for the corresponding brane positions. Before moving further, we would like to mention that we have considered an intermediate point of radial coordinate $r$ between the early and late time brane positions. If we denote this point as $r_t$, then in the early time $r<<r_t$ and in the late time $r>>r_t$. All of our analyses are done in the very early and very late times of the braneworld universe.
\subsubsection*{Pure AdS}
In the absence of any matter source in our universe, the Israel junction condition in eq.\eqref{diff_brane_position} becomes
\begin{equation}
    \paren{\frac{dr}{d\tau}}^2=\paren{\frac{\scrT^2}{36}-\frac{1}{R^2}}r^2~.
\end{equation}
The solution of the above differential equation gives the time-dependent radial position of the brane for an eternally inflating universe, which reads
\begin{equation}
    r=r_ie^{H\tau}
\end{equation}
where $r_i$ is the brane position at $\tau=0$ and $H$ is the Hubble parameter. The Hubble parameter here is given by $\sqrt{\frac{\scrT^2}{36}-\frac{1}{R^2}}$. \\
Now we can transform this expression for the brane position in terms of the inverse radial coordinate $z$. The time-dependent brane position in the $z$ coordinate is given by
\begin{equation}\label{zbar_exact}
    \z=\frac{1}{r_i}e^{-H\tau}~.
\end{equation}
In the early time of this braneworld model, we have $H\tau<<1$. In this limit, the brane position becomes 
\begin{equation}\label{early brane pos AdS}
    \z\simeq\frac{1}{r_i}\paren{1-H\tau}~.
\end{equation}
Here, early time refers to the time at which the position of the brane is at $r=r_i$. It should be noted that for the dS branch, $\tau \to -\infty$ (and not $\tau \to 0$) corresponds to the asymptotic past. Hence, the expansion of eq.\eqref{zbar_exact} around $\tau\to 0$ to get eq.\eqref{early brane pos AdS} is only a local expansion around a finite scale factor. Thus, the terminology “early time” mentioned here refers to early evolution with respect to the chosen initial slice $r=r_i$ corresponding to the brane position, and not to the asymptotic past limit $\tau\to-\infty$ .\\
The late time is denoted by $\tau\to \infty$. In this limit, the brane position becomes
\begin{equation}\label{late brane pos AdS}
    \z(\tau)\approx\frac{e^{-H\tau}}{r_0}~.
\end{equation}
These early and late time brane positions will later be helpful to determine the time-dependent holographic entanglement entropy and volume complexity for an eternally inflating universe.
\subsubsection*{Radiation and matter dominant universe }
To find out the brane position in terms of time, we have to start from eq.\eqref{diff_brane_position} and choose $p=1$. This gives 
\begin{equation}
    \paren{\frac{dr}{d\tau}}^2=\paren{\frac{\scrT^2}{36}-\frac{1}{R^2}}r^2+\frac{m}{r^2R^2}+\frac{\rc}{R^2r}~,
\end{equation}
where we have defined $\rc=\rho_1$, and $\rho_1$ is the energy density of the 1-brane string gas in the bulk.\\
Now, for finding out the non-trivial contribution from matter and radiation, we have to solve for $\scrT=\scrT_c\equiv\frac{6}{R}$, which in turn gives
\begin{equation}
     \paren{\frac{dr}{d\tau}}^2=\frac{m}{r^2R^2}+\frac{\rc}{R^2r}=\frac{\m}{r^2}+\frac{\T\rc}{r}~.
\end{equation}
It would be worthy to mention that when $\scrT\neq\scrT_c$, one gets three component (radiation, matter, dark energy) FLRW universe where the effect of dark energy comes through the vacuum energy term $\sqrt{\frac{\scrT^2}{36}-\frac{1}{R^2}}$ \cite{Sahni:2002dx,Binetruy:1999ut}. The primary reason for working with the critical brane tension is to focus solely on the effects of radiation, matter, and exotic matter, without accounting for the contribution of dark energy in the background. \\
In the early time limit ($\tau\rightarrow0$), we have to remember that the initial brane position (that is, the brane position at $\tau=0$) is $r_i$, and in this limit, after integrating the above differential equation, we get with $r_t \equiv \frac{\m}{\T\rc}$
\begin{equation}\label{eq:61}
    \tau=\int_{r_i}^{r}dr\frac{r }{\sqrt{\m+\T\rc\;r}}~.
\end{equation}
\begin{equation}\label{eq:104}
    \paren{\frac{r^2}{2}-\frac{r^2_i}{2}}-\paren{\frac{r^3}{6r_t}-\frac{r_i^3}{6r_t}}=\tau\sqrt{\m}~.
\end{equation}
Clearly, the time has come as a function of brane position, but in the later sections, we need the brane position as a function of time. Upon inverting this perturbatively, we get 
\begin{equation}
    r\simeq r_i+\frac{\sqrt{\m}\;\tau}{r_i}+\frac{\sqrt{\m}\;\tau}{2r_t}~.
\end{equation}
Now defining $\z\equiv\frac{1}{r}$, $z_i\equiv\frac{1}{r_i}$,$z_t\equiv\frac{1}{r_t}$ we get
\begin{equation}\label{early_brane_rad+matt}
    \z\simeq z_i\paren{1-\sqrt{\m}\;\tau\paren{z_i^2+\frac{z_iz_t}{2}}}~.
\end{equation}
Again for the late time ($\tau\rightarrow\infty$) we can write 
\begin{equation}
    \tau=\int_{r_i}^{r_t}dr\frac{r }{\sqrt{\m+\T\rc\;r}}+\int_{r_t}^{r} d\T r\frac{\sqrt{\T r}}{\sqrt{\T\rc}}\paren{1-\frac{r_t}{2\T r}}~.
\end{equation}
Again, this is the time that arises as a function of brane position; we must invert this perturbatively to obtain the brane position as a function of time. This gives
\begin{equation}
    r\simeq\paren{\frac{3}{2}}^{2/3}\T\rc^{1/3}\tau^{2/3}\Bigg(1-\paren{\frac{2}{3}}^{2/3} \frac{r_t}{\T\rc^{1/3}}\tau^{-2/3}+\frac{2}{3\sqrt{\m}}\frac{1}{\tau}\paren{\frac{2r_t^2}{3}-\frac{r_i^2}{2}+\frac{r_i^3}{6r_t}}\Bigg)~.
\end{equation}
Which can be recast similar to eq.\eqref{early_brane_rad+matt} as 
\begin{equation}\label{late_brane_rad+matt}
    \z \simeq \paren{\frac{2}{3}}^{2/3}\T\rc^{-1/3}\tau^{-2/3}\Bigg(1+\paren{\frac{2}{3}}^{2/3} \frac{r_t}{\T\rc^{1/3}}\tau^{-2/3}-\frac{2\sqrt{z_t}}{3\sqrt{\T\rc}}\frac{1}{\tau}\paren{\frac{2}{3z_t^2}-\frac{1}{2z_i^2}+\frac{z_t}{6z_i^3}}\Bigg)~.
\end{equation}
It is worth noting one thing that as we are doing the perturbation in the late time so one can not demand that $r\rightarrow r_i$ for $\tau \rightarrow 0$ because $\tau \rightarrow 0$.
\subsection*{Radiation and exotic matter dominant universe}
We will now find out the time-dependent brane position for a universe with coexisting radiation and exotic matter. Just like the universe with radiation and matter, here also, we will find the early and late time brane positions. To start with, we will take the junction condition in eq.\eqref{diff_brane_position} along with the lapse function of the black-brane in eq.\eqref{lapse fn total}. As we are dealing with a universe with co-existing radiation and exotic matter, we would take $p=2$ for the lapse function in eq.\eqref{lapse fn total}. This choice of lapse function, along with eq.\eqref{diff_brane_position} gives
\begin{equation}\label{int eq rad+ex}
    \paren{\frac{dr}{d\tau}}^2=\frac{\m}{r^2}+\chi = \frac{\T m}{r^2}\left(1+\frac{r^2}{r_t^2}\right)~.
\end{equation}
Here we have defined $r_t=\sqrt{\frac{m}{\chi}}$.
In the early time, that is when $r<< \T r_t$, the above equation can be solved by solving the following integral 
\begin{equation}
    \int_{r_i} ^ r dr\frac{r}{\sqrt{m}}\paren{1-\frac{r^2}{2{\T r_t}^2}}\simeq \tau~.
\end{equation}
The above equation can be easily solved, and a series inversion at the leading order can be done to obtain an expression for the brane's radial position $r$ as a function of the cosmological time $\tau$. This reads
\begin{equation}
    r=r_i\Bigg[1+\sqrt{\m}\paren{\frac{1}{r_i^2}-\frac{1}{2\T r_t^2}}\tau\Bigg]~.
\end{equation}
We can transform the above relation to get the time-dependent brane position in inverse radial coordinate $z$. Keeping terms up to the first order in $\tau$, we get
\begin{equation}\label{early brane rad+ex}
    \z=z_i\Bigg[1-\sqrt{\m}\paren{{z_i^2}-\frac{\T z_t^2}{2}}\tau\Bigg]~.
\end{equation}
Similarly, we can derive an expression for the time-dependent brane position in the late time. At a late time, we will deal with a radial position of the brane in the regime $r>> \T r_t$. In this regime, we need to break the integral in two parts, one from $r_i$ to $\T r_t$ and another from $\T r_t$ to $r$. Therefore, we can further write eq.\eqref{int eq rad+ex} in the following integral form
\begin{equation}
    \frac{1}{\sqrt{\m}}\int_{r_i}^{\T{r_t}}dr \paren{1-\frac{r^2}{2\T{r_t}^2}}+\frac{1}{\sqrt{\T{\chi}}}\int_{\T{r_t}}^{r}d\T{r}\paren{1-\frac{\T{r_t}^2}{2 \T{r}^2}}\simeq \tau~.
\end{equation}
Just like the early time scenario, the above integral can be very easily solved, and a series inversion in the leading order in $\tau$ gives the brane position $\z$ as a function of the cosmological time $\tau$, which reads
\begin{equation}\label{late brane rad+ex}
    \z\simeq\frac{1}{\sqrt{\chi}\tau}\Bigg[1-\frac{1}{\sqrt{\m}}\paren{\frac{3}{8\T z_t^2}-\frac{1}{2z_i^2}}\frac{1}{\tau}-\frac{1}{2\chi\T z_t^2 \tau^2}\Bigg]~.
\end{equation}
The above brane position for the universe with co-existing radiation and exotic matter clearly tells that in the late time the brane position changes as $\frac{1}{\tau}$ in the leading order. This suggests that in the late time of this kind of universe, the effect of exotic matter density is dominant over radiation density. \\
These time-dependent brane positions (for early and late times) will later be useful to compute the time-dependent holographic entanglement entropy and holographic volume complexity of different kinds of universes, like universes with co-existing radiation-dark matter and radiation-exotic matter.
\section{Holographic entanglement entropy}
In this section, we will compute the time-dependent entanglement entropy for a universe with co-existing matter sources like radiation-dark matter and radiation-exotic matter in a holographic manner. To calculate the time-dependent holographic entanglement entropy of our universe, we will mainly use the Ryu-Takayanagi (RT) formula and the Israel junction conditions for different types of universes. Before proceeding further, we will discuss about the Ryu-Takayanagi formula to compute the holographic entanglement entropy. The RT formula relates the entanglement entropy of a strongly coupled CFT with the area of a codimension-two RT surface, which is extended in the bulk. If we consider a subsystem of finite size for the strongly interacting CFT on the $d$-dimensional boundary, the corresponding holographic entanglement entropy is given by the RT formula. This states that the entanglement entropy of a subsystem $A$ is given by
\begin{equation}
    S_{HEE}=\frac{1}{4G_{N}^{(d+1)}}Area(\Gamma^{A}_{min})
\end{equation}
where $G_{N}^{(d+1)}$ is the Newton's gravitational constant in $d+1$-dimensional bulk spacetime. $\Gamma^{A}_{min}$ denotes the $d-1$-dimensional static minimal surface, such that $\partial\Gamma^{A}_{min}=\partial A$.\\
In the upcoming subsections, we will use this formula for a circular subsystem of radius $l$ and compute the holographic entanglement entropy of some part of the universe with dimension $l$ with the rest of the universe.
\subsection{ Universe with no matter}\label{subsection Ads entropy}
In the braneworld model, to calculate the HEE, we have to consider pure
AdS$_5$ spacetime in the bulk, for the universe with no matter. In order to do so, we will consider a static brane. However, one should remember that this model does not
explain the expansion of the universe, but it is useful for understanding the time
evolution of HEE in various expanding universes. The spacetime metric for $AdS_{5}$
spacetime in the absence of any matter source in the universe is given by \cite{Paul:2025gpk}
\begin{equation}
ds^2 = \frac{R^2}{z^2} \Bigg( dz^2 - dt^2 + du^2 + u^2\, d\Omega_2^2 \Bigg)
\end{equation}
after reducing the metric and parameterisation of `z' as a function of `u',
We get the area functional for pure AdS spacetime
\begin{align}\label{Ads int}
A^{(\text{AdS})} = \Omega_2 \int_{0}^{l} du \, \frac{u^{2}\sqrt{1 + z'^2}}{z^3} \, .
\end{align}
 We identify the integrand of the above area functional to be a Lagrangian of the form $\mathcal{L} = \mathcal{L}(z, z')$. Hence, the Euler-Lagrange equation of motion corresponding to this Lagrangian reads
\begin{align}\label{EL_AdS}
     3u +3uz'^{2}(u) + uz(u)z''(u) + 2z(u)z'(u) + 2z(u)z'^{3}(u) = 0 .
\end{align}
 To evaluate the form of the RT surface, we need to solve the above differential equation
 equation. After a careful analysis, we can write the above
 equation in the following form
 \begin{align}
     u\,\frac{d}{du}\big( z(u)z'(u) + u \big) + 2\big(1 + z'^2(u)\big)\big(z(u)z'(u) + u\big) = 0.
 \end{align}
  One can see that the above equation has a trivial solution of the form $ z(u)z'(u) + u =0$. Hence, the solution of z(u) is given by
  \begin{equation}
z(u) = \sqrt{c_1 - (c_2 + u)^{2}}
\end{equation}
where $c_1$ and $c_2$ are arbitrary integration constants. To determine the value of $c_1$
and $c_2$, we need to consider the boundary conditions for $z(u)$. The smoothness condition of the RT surface requires that the surface must be regular, that is, it should not have any cusp, sharp corner, or discontinuity, at the turning point, which implies that the first derivative of the 
function vanishes at $u = 0$, leading to $z'(0) = 0$ due to the rotational symmetry. Now this
condition fixes the integration constant $c_2 = 0$. Another condition that we apply is
that $z(l) = \bar{z}$, where $\bar{z}$ is the position of the brane. This condition leads to
$c_1 = l^2 + \bar{z}^2$.
Therefore, the final form of the RT surface has the form
\begin{align}\label{RT surface profile AdS}
z(u) = \sqrt{l^2 + \bar{z}^2 - u^2}.
\end{align}
We would like to mention that inside the subsystem, on the brane, $l^2 + \bar{z}^2$ is always
much less than $u^2$. Solving the integral in eq.\eqref{Ads int}, we get
\begin{equation}
A^{(AdS)}=\frac{\Omega_{2}}{2} \left[
\frac{l\,\sqrt{l^2 + \bar{z}^2}}{\bar{z}^2}
-
\operatorname{arctanh}\!\left(\frac{l}{\sqrt{l^2 + \bar{z}^2}}\right)
\right]~.
\end{equation}
Expanding the argument under $tanh^{-1}$ up to the first order and with some algebra, we get the expression for the minimal area of pure AdS spacetime to be
\begin{align}
    A^{(AdS)}\approx\frac{\Omega_{2}}{2}\Bigg(\frac{l^3}{\sqrt{l^2+\z^2}\z^2}\Bigg)~.
\end{align}
Therefore, from the RT formula, we can write the holographic entanglement entropy as
\begin{equation}\label{HEE AdS gen}
    S_{HEE}^{(AdS)}\approx\frac{\Omega_{2}}{8\pi G_5}\Bigg(\frac{l^3}{\sqrt{l^2+\z^2}\z^2}\Bigg)~.
\end{equation}
Now to know how this area functional grows with time in the early time limit, we have to use eq.\eqref{early brane pos AdS}. Upon substituting the time-dependent radial position of the brane for an eternally inflating universe from eq.\eqref{early brane pos AdS} in the early time, we get the following expression of HEE
\begin{equation}
    S^{(AdS)}_{HEE,early}=\frac{\Omega_2}{4G_{5}}\left[\frac{lr_{0}^3 (1+2H\tau)}{\sqrt{r_{0}^2 l^2+1}}\paren{1+\frac{H\tau}{r_{0}^2 l^2+1}}\right]~.
\end{equation}
Similarly, in order to obtain the late-time behaviour of the HEE for an eternally inflating universe, we need to use eq.\eqref{late brane pos AdS}. Upon substituting the expression of $\z$ from eq.\eqref{late brane pos AdS} in eq.\eqref{HEE AdS gen}, we get
\begin{equation}
    S^{(AdS)}_{HEE,late}=\frac{\Omega_2}{4 G_{5}}\left[\frac{r_{0}^2 e^{2H\tau}}{2}\paren{1-\frac{e^{-2H\tau}}{2l^2 r_{0}^2}}+\frac{1}{2l^2}\paren{1-\frac{3e^{-2H\tau}}{l^2 r_{0}^2}}\right]~.
\end{equation}
From the above expression of HEE, it is clear that at a late time, HEE changes as $e^{2H\tau}$. An important point to note is that as the brane moves, the length of the subsystem on the brane is not the actual length of the system. This length is called the comoving length of the subsystem, that is, $l$. The actual length measured by an observer who is sitting at rest outside the comoving frame is given by
\begin{equation}\label{L orginial}
    L=\frac{R}{\z (\tau)}l~.
\end{equation}
Therefore, in the late time of an eternally inflating universe, the actual length of the circular subsystem is proportional to $e^{H\tau}$. Thus, the HEE in the late time of an eternally inflating universe follows an area law, which is expected in the present time.
\subsection{Universe with coexisting radiation and matter}
According to the thermal history of our universe \cite{WMAP:2010qai,WMAP:2010sfg,Planck:2014loa,Planck:2018vyg}, the existence of a radiation and dark matter-dominated universe is well established. In this section, we are going to see how the entanglement entropy of such a universe evolved in the early time when there was radiation domination and how the subdominant matter component affected the dynamics there. Similarly, we shall check the late time case where there is matter domination, and radiation is a subdominant component. The area functional for both matter and radiation reads
\begin{equation}\label{RT_area_rad_matt}
    A^{\textit{rad+mat}}=\Omega_{2}\int_0^ldu\frac{u^2}{z(u)^3}\sqrt{1+\frac{z'^2}{f_{m,r}\left(z(u)\right)}}~.
\end{equation}
The blackening factor we have taken is $f_{m,r}(z(u))=1-\T{\rc} z(u)^3-\T{m}z(u)^4$ with $\T{m}=\frac{m}{R^8}$ and $\T\rc=\frac{\rc}{R^6}$.
Now, the integrand of the  area function is mathematically a Lagrangian-like function, and the Euler-Lagrange (EL) equation reads 
\begin{align}
&6u\T\rc^{2} z(u)^{6}
+ 12 \m \T\rc u  z(u)^{7}
+ 6 \m^{2} u z(u)^{8}
+ 6u \left(1 + \left(z'(u)\right)^{2}\right) - 3u \T\rc z(u)^{3} \left(4 + \left(z'(u)\right)^{2}\right)
\nonumber \\
&\quad - 2\m z(u)^{5} \left(2 z'(u) + u z''(u)\right) + 2 z(u) \left(2 \left(z'(u) + \left(z'(u)\right)^{3}\right) + u z''(u)\right)
\nonumber\\
&\quad - 2 z(u)^{4} \left(6 \m u + 2 \T\rc z'(u) + \m u \left(z'(u)\right)^{2} + u \T\rc z''(u)\right)=0~.
\end{align}
As we are interested in solving this perturbatively, we are going to drop $\mathcal{O}\paren{\T\rc^2}, \mathcal{O}\paren{\m^2}$ and $\mathcal{O}\paren{\m\T\rc }$, which reads
\begin{align}
&6u \left(1 + \left(z'(u)\right)^{2}\right) - 3u \T\rc z(u)^{3} \left(4 + \left(z'(u)\right)^{2}\right)
\nonumber \\
&\quad - 2\m z(u)^{5} \left(2 z'(u) + u z''(u)\right) + 2 z(u) \left(2 \left(z'(u) + \left(z'(u)\right)^{3}\right) + u z''(u)\right)
\nonumber\\
&\quad - 2 z(u)^{4} \left(6 \m u + 2 \T\rc z'(u) + \m u \left(z'(u)\right)^{2} + u \T\rc z''(u)\right)=0~.
\end{align}

To solve the above equation, we are taking the perturbative form of $z(u)$ as
\begin{equation}
    z(u)=z_0(u)+\m z_1(u)+\T{\rc} z_2(u)~.
\end{equation} 
We get three EL equations corresponding to various powers of the perturbation parameters shown below $\mathcal{O}(\T\rc^{0},\m^0)$,$\mathcal{O}(\T\rc^{0},\m^1)$ and $\mathcal{O}(\T\rc^{1},\m^0)$ as follows\\ \\
\textbf{Terms of $\mathcal{O}(\T\rc^{0},\m^0)$:}

\begin{equation}
\left( 3 u + 2 z_0(u) z_0'(u) \right) \left( 1 + (z_0'^2(u) \right) + 
 u z_0(u) z_0''(u)=0~.
\end{equation}\\
\textbf{Terms of $\mathcal{O}(\T\rc^{0},\m^1)$:}
\begin{align}\label{matt_rad_diff_z1}
& u z_1(u) z_0''(u) + u z_0(u) z_1''(u)- 
u z_0(u)^5 z_0''(u)-6 u z_0(u)^4 - 2 z_0(u)^5 z_0'(u) \nonumber\\&+ 
2 z_1(u) z_0'(u) - u z_0(u)^4 (z_0'(u))^2 + 
2 z_1(u) (z_0'(u))^3 + 2 z_0(u) z_1'(u) \nonumber\\&+ 
6 u z_0'(u) z_1'(u) +6 z_0(u) (z_0'(u))^2 z_1'(u) =0~.
\end{align}\\
\textbf{Terms of $\mathcal{O}(\T\rc^{1},\m^0)$:}
\begin{align}\label{matt_rad_diff_z2}
    & 2z_0(u)\paren{\paren{2+6z_0'^{2}(u)}z_2'(u)+uz_2''(u)}-2z_0^4(u)\paren{2z_0'(u)+uz_0''(u)}+2uz_2(u)z_0''(u)\nonumber \\&+4z_0'(u)\paren{z_2(u)\paren{1+z_0^{'2}(u)}+3uz_2^{'}(u)}-3uz_0^3(u)\paren{4+z_0^{'2}(u)}=0~. 
\end{align}
So clearly we can see from the zeroth order equation that it is exactly the EL eq.\eqref{EL_AdS} corresponding to the AdS space and the solution reads the same as earlier
\begin{equation}
    z_0(u) = \sqrt{l^2 + \bar{z}^2 - u^2}.
\end{equation}
Now the job remains to find the explicit solution of $z_1(u)$ and $z_2(u)$. To do this, we will proceed with the help of a series solution, but before that, we need to put the form of $z_0(u)$ into eq.\eqref{matt_rad_diff_z1} and eq.\eqref{matt_rad_diff_z2}. After this substitution, we can easily perform a series solution corresponding to $z_1 (u)$ and $z_2 (u)$. After some calculations, we arrive at the following forms of $z_1 (u)$ and $z_2 (u)$
\begin{equation}
\begin{split}
    z_{1}(u)&=a_{01}\paren{1+\frac{u^2}{2\C}+\frac{3u^4}{8\C^2}+\mathcal{O}\paren{\frac{u^6}{\C^3}}}\\ &+\C^{5/2}\left(\frac{u^2}{\C}\right)\left(\frac{1}{2}-\frac{13}{40}\frac{u^2}{\C}+\frac{23}{560}\frac{u^4}{\C^2}\right. \\& \left.+\mathcal{O}\paren{\frac{u^5}{\C^{5/2}}}\right)~
\end{split}
\end{equation}
\begin{align}
    z_2(u)=&a_{02}\Bigg[1+\frac{u^2}{2\C}+\frac{3}{8}\frac{u^4}{\C^2}+\mathcal{O}\paren{\frac{u^6}{\C^3}}\Bigg]\nonumber \\
    &+\C^2\Bigg[\frac{u^2}{2\C}-\frac{u^4}{\C^2}+\mathcal{O}\paren{\frac{u^6}{\C^3}}\Bigg]
\end{align}
where $a_{01}$ and $a_{02}$ are constants to be determined from the boundary conditions. \\ The boundary conditions demand that $z_1(l)=0=z_2(l)$ and $z_0(l)=\z$. Applying these boundary conditions, we get the forms of $z_1(u)$ and $z_2(u)$ as 
\begin{align}
    &z_1(u)=\C^{\frac{5}{2}}\paren{\frac{-l^2}{2\C}+\frac{u^2}{2\C}+\frac{2}{5}\frac{l^4}{\C^2}-\frac{l^2u^2}{4\C^2}-\frac{3}{20}\frac{u^4}{\C^2}} \\
    &z_2(u)=\C^2\paren{-\frac{l^2}{2\C}+\frac{u^2}{2\C}+\frac{3}{10}\frac{l^4}{\C^2}-\frac{u^4}{20\C^2}-\frac{l^2u^2}{4\C^2}}~.
\end{align}
With all these in hand, we proceed to calculate one of our main goals, calculating the RT area function, which we get by using the form of $z_0(u),z_1(u)$ and $z_2(u)$ and putting those into eq.\eqref{RT_area_rad_matt}, we get the perturbative form of the RT surface as
\begin{align}
    A^{rad+mat}=\Omega_2\Bigg[&\frac{l^3}{2\z^2\sqrt{\C}}+\m\C^2\paren{\frac{1}{10}\frac{l^5}{\C^\frac{5}{2}}} \nonumber \\&+\rc \C^\frac{3}{2}\paren{\frac{1}{10}\frac{l^5}{\C^\frac{5}{2}}}\Bigg]~
\end{align}
which in turn gives the HEE for a radiation and matter dominated universe in terms of the brane position $\z$ as
\begin{align}\label{entropy raw rad+matt}
    S^{rad+mat}=\frac{\Omega_2}{4 G_{5}}\Bigg[&\frac{l^3}{2\z^2\sqrt{\C}}+\m\C^2\paren{\frac{1}{10}\frac{l^5}{\C^\frac{5}{2}}} \nonumber \\& +\rc \C^\frac{3}{2}\paren{\frac{1}{10}\frac{l^5}{\C^\frac{5}{2}}}\Bigg]~.
\end{align}
Finding out how the entropy scaled in the early time is not a difficult job, as using eq.\eqref{early_brane_rad+matt} and inserting into the above equation, we get the time-dependent HEE as
\begin{align}
    S&^{rad+matt}_{early}=\frac{\Omega_2}{4 G_{5}}\frac{l^5}{5}\Bigg[\frac{5}{2l^2z_i^2\sqrt{l^2+z_i^2}}\Bigg(1+\m\frac{5z_i^2+z_iz_t+2l^2}{l^2+z_i^2}\tau\Bigg)\nonumber\\&+\frac{20}{\ci^\frac{5}{2}}\paren{\G\tau}-\frac{2\m}{\sqrt{\ci}}\paren{1+2\G\tau}\nonumber\\&-\frac{2\T\rc}{\ci}\paren{1+4\G\tau}\Bigg]+\frac{l^3}{3}\Bigg[\frac{1}{\ci^\frac{3}{2}}\paren{6\G\tau}\nonumber \\& +\frac{3l^2\m}{2\sqrt{\ci}}\paren{1+2\G\tau}+\frac{3l^2\T\rc}{2\ci}\paren{1+4\G\tau}\Bigg]~.
\end{align}
The above expression clearly shows that in the early times of a universe containing both radiation and matter. The holographic entanglement entropy scales as $\tau$ in the leading order. An important thing to note is that in the early times, the leading order of HEE and the time-dependent brane position scale in the same way as the cosmological time. We can also see that due to the presence of two coexisting matter components, the result of time-dependent HEE contains energy densities corresponding to both radiation and dark matter.\\
Now we will move forward to compute the late time behaviour of the HEE with the cosmological time $\tau$. To obtain a late-time expression of the HEE for a universe with coexisting radiation and matter, we will put the time-dependent brane position from eq.\eqref{late_brane_rad+matt} in eq.\eqref{entropy raw rad+matt}, which gives 
\begin{align}
   S^{rad+matt}_{late}=&\frac{3}{4}\,\paren{\frac{3}{2}}^{1/3} \, l^{2}\, \T\rc^{2/3}\, \tau^{4/3}
\;-\;
\paren{\frac{3}{2}} ^{2/3}\frac{1}{z_t} \, l^{2}\, \T\rc^{1/3}\, \tau^{2/3}\nonumber \\ &
-
\frac{2 \paren{\frac{2}{3}}^{2/3} l^{2}\, }{ \T\rc^{1/3} z_t^{3}\tau^{2/3}}
\;+\;
\frac{1}{20}\!\left( -5 + 2 l^{3} \T\rc + 2 l^{4} \frac{\T\rc}{z_t} + \frac{30 l^{2}}{z_t^2} \right)
~.
\end{align}
From the above expression of the HEE in the late time regime of this universe with radiation and dark matter, the leading order term is of the order $\tau^{4/3}$. The sub-leading term is of the order $\tau^{2/3}$. As we already know, the physical length scale in this case is proportional to $\tau^{2/3}$ in the leading order; our result is also consistent with the area law in the leading order. Another point is to note that the leading order term shows the dominant contribution from the dark matter energy density, and contributions of radiation density are subdominant. This observation is consistent with the thermal history of the universe \cite{WMAP:2010qai,WMAP:2010sfg,Planck:2014loa,Planck:2018vyg}, where we know that in the early times, the universe was dominated by radiation and in the late times, it has dark matter dominance.
\subsection{Universe with coexisting radiation and matter at the intermediate regime}\label{sub sec 4.3}
In this subsection, we discuss the coexistence of the two components, namely, 
radiation and matter, in the intermediate regime where 
$\tilde{m}/r^4 \sim \rho_p /r^{3}$.
In order to investigate the intermediate regime, at first we need to evaluate eq.\eqref{eq:61}, which reads
\begin{align}\label{integral exact}
    \tau&=\int_{r_i}^{r}\frac{r dr}{\sqrt{\m+\T\rc~r}}\nonumber \\ &=\frac{2}{3\T\rc^2}\Bigg[(-2\m+\T\rc~ r)\sqrt{\m+\T\rc~ r}+\T C\Bigg]
\end{align}
where $\T C= (2\m-\T\rc~ r_i)\sqrt{\m+\T\rc~ r_i}$~.\\
As we are in the intermediate regime, the radial brane position $r$ is close to the matter-radiation equality point $r_t$. Hence, we substitute $r=r_t(1+\epsilon)$ (with $\epsilon<<1$) in eq.\eqref{integral exact} and find the solution of $\epsilon$ in terms of the cosmological time ($\tau$) by dropping the $\mathcal{O}(\epsilon^2)$ terms. This gives
\begin{equation}\label{epsilon1}
    \epsilon=\frac{1}{3}\Bigg(\frac{2}{\m}\Bigg)^{\frac{3}{2}}\Bigg[\frac{3}{2}\T\rc^2\tau - \T C\Bigg]+\frac{4}{3}~.
\end{equation}
In the inverse radial coordinate $\z=\frac{1}{r}$, and using the definition $z_t=\frac{1}{r_t}$ the brane position can be written as 
\begin{equation}\label{z bar inter}
    \z=z_t(1-\epsilon)~.
\end{equation}
Substituting the above expression for $\z$ into the form of the HEE in eq.\eqref{entropy raw rad+matt}, we obtain up to $\mathcal{O}(\epsilon)$ 
\begin{align}
     S^{rad+mat}_{int}&\sim\frac{\Omega_2}{4G_5}\Bigg[\frac{l^3}{2z_t^2\big(l^2+z_t^2\big)^{\frac{1}{2}}}\big(1+2\epsilon\big)\Bigg]~.
\end{align}
Since the actual length of the subsystem $L$ is given by $L\sim\frac{l}{z_t}(1+\epsilon)$ (eq.\eqref{L orginial}), we observe that the HEE scales as $L^2$, which is the area law.\\
Finally, using the form of $\epsilon$ from eq.\eqref{epsilon1}, we obtain the HEE (in the intermediate regime where both radiation and matter play an important role) in terms of $\tau$ to be
\begin{align}
     S^{rad+mat}_{int}&\sim\frac{\Omega_2}{4G_{5}}\Bigg[\frac{l^3}{2z_t^2\big(l^2+z_t^2\big)^{\frac{1}{2}}}\Bigg\{1+2\Bigg(\frac{1}{3}\Big(\frac{2}{\m}\Big)^{\frac{3}{2}}\Big(\frac{3}{2}\T\rc^2\tau - \T C\Big)+\frac{4}{3}\Bigg)
     \Bigg\}
     \Bigg]~.
\end{align}
\subsection{Universe with coexisting radiation and exotic matter}
In this subsection, we do the detailed calculation for the time-dependent holographic entanglement entropy for a universe with coexisting radiation and some exotic matter. In order to do so, we will use the black brane lapse function of eq.\eqref{lapse fn total} and set $p=2$. This will help us to study a four-dimensional FLRW universe on the brane with coexisting radiation and exotic matter. The area functional in this case is given by
\begin{equation}\label{RT_area_rad_ex}
    A^{\textit{rad+ex}}=\Omega_{2}\int_0^ldu\frac{u^2}{z(u)^3}\sqrt{1+\frac{z'^2(u)}{f_{r,e}\left(z(u)\right)}}~.
\end{equation}
Here, $f_{r,e}$ is the lapse function corresponding to $p=2$ in eq.\eqref{lapse fn total} and is given by 
\begin{equation}
    f_{r,e}=1-\T\chi z(u)^2-\m z(u)^4
\end{equation}
where $\T \chi =\frac{\chi}{R^4}$ and $\m =\frac{m}{R^8}$. \\
The Euler-Lagrange (EL) equation corresponding to the above area functional (in eq.\eqref{RT_area_rad_ex}) reads to the following differential equation
\begin{align}
    &6 \m\T\chi u  z(u)^6 + 3 \m^2 u z(u)^8 + 3 u \left(1 + \left(\frac{d z(u)}{d u}\right)^2\right) - 2 u \T\chi z(u)^2 \left(3 + \left(\frac{d z(u)}{d u}\right)^2\right)\nonumber \\ &- u z(u)^4 \left(6 \m - 3 \T\chi^2 + \m \left(\frac{d z(u)}{d u}\right)^2\right) - \T\chi z(u)^3 \left(2 \frac{d z(u)}{d u} + u \frac{d^2 z(u)}{d u^2}\right)\nonumber\\& - \m z(u)^5 \left(2 \frac{d z(u)}{d u} + u \frac{d^2 z(u)}{d u^2}\right)+ z(u) \left(2 \left(\frac{d z(u)}{d u} + \left(\frac{d z(u)}{d u}\right)^3\right) + u \frac{d^2 z(u)}{d u^2}\right)=0~.
\end{align}
Now, similar to the previous section, we can keep terms up to the first order of $\m$, $\T\chi$. Therefore we can simply drop terms of $\mathcal{O}(\T\chi^2)$, $\mathcal{O}(\m^2)$ and $\mathcal{O}(\m\T\chi)$. Thus, the above non-linear differential equation simplifies to the following expression
\begin{align}
    &3 u \left(1 + \left(\frac{d z(u)}{d u}\right)^2\right) - 2 u \T\chi z(u)^2 \left(3 + \left(\frac{d z(u)}{d u}\right)^2\right)-\m u z(u)^4 \left(6  +  \left(\frac{d z(u)}{d u}\right)^2\right) \nonumber\\ & - \T\chi z(u)^3 \left(2 \frac{d z(u)}{d u} + u \frac{d^2 z(u)}{d u^2}\right)- \m z(u)^5 \left(2 \frac{d z(u)}{d u} + u \frac{d^2 z(u)}{d u^2}\right)\nonumber\\& + z(u) \left(2 \left(\frac{d z(u)}{d u} + \left(\frac{d z(u)}{d u}\right)^3\right) + u \frac{d^2 z(u)}{d u^2}\right)=0~.
\end{align}
In order to determine the profile of the RT surface, we need to solve the above non-linear differential equation. We will use the same methodology just like the previous subsection. Let us consider the RT surface profile is given by
\begin{equation}\label{RT profile rad+ex}
    z(u)=z_0(u)+\m z_1(u) +\T\chi z_3(u)
\end{equation}
where $z_0(u)$ is the profile of the unperturbed RT surface for a universe with no matter sources, $z_1(u)$ and $z_3(u)$ are the corrections to the RT surface profile due to the presence of radiation and exotic matter, respectively.
By putting the form of the RT surface from eq.\eqref{RT profile rad+ex} in the above equation, it is obvious to say that, we can get three EL equations corresponding to various powers of the perturbation parameters. If we collect the powers of $\mathcal{O}(\T\chi^{0},\m^0)$,$\mathcal{O}(\T\chi^{0},\m^1)$ and $\mathcal{O}(\T\chi^{1},\m^0)$, we will get three differential equations gives as follows.\\ \\
\textbf{Terms of $\mathcal{O}(\T\rc^{0},\m^0)$:}

\begin{equation}\label{diff eq z0 rad+ex}
\left( 3 u + 2 z_0(u) z_0'(u) \right) \left( 1 + (z_0'^2(u) \right) + 
 u z_0(u) z_0''(u)=0~.
\end{equation}\\
\textbf{Terms of $\mathcal{O}(\T\rc^{0},\m^1)$:}
\begin{align}\label{matt_ex_diff_z1}
& u z_1(u) z_0''(u) + u z_0(u) z_1''(u)- 
u z_0(u)^5 z_0''(u)-6 u z_0(u)^4 - 2 z_0(u)^5 z_0'(u) \nonumber\\&+ 
2 z_1(u) z_0'(u) - u z_0(u)^4 (z_0'(u))^2 + 
2 z_1(u) (z_0'(u))^3 + 2 z_0(u) z_1'(u) \nonumber\\&+ 
6 u z_0'(u) z_1'(u) +6 z_0(u) (z_0'(u))^2 z_1'(u) =0~.
\end{align}\\
\textbf{Terms of $\mathcal{O}(\T\chi^{1},\m^0)$:}
\begin{align}\label{diff eq z3 rad+ex}
    &
u z_3(u) z_0''(u) 
- z_0(u)^3 \left( 2 z_0'(u) + u z_0''(u) \right) 
+ z_0(u) \left( (2 + 6 (z_0'(u))^2) z_3'(u) + u z_3''(u) \right)\nonumber\\&-2 u z_0(u)^2 \left( 3 + (z_0'(u))^2 \right) 
+ 2 z_0'(u) \left( z_3(u) \left( 1 + (z_0'(u))^2 \right) + 3 u z_3'(u) \right) =0~.
\end{align}
Now we will solve the above set of differential equations. As shown earlier in the subsection \eqref{subsection Ads entropy}, that eq.\eqref{diff eq z0 rad+ex} has a trivial solution of the form as follows
\begin{equation}
    z_0(u)=\sqrt{\C-u^2}~.
\end{equation}
The other two equations \eqref{matt_ex_diff_z1} and \eqref{diff eq z3 rad+ex} can be solved perturbatively and using a series solution method like the previous subsection. Keeping terms up to $\mathcal{O}(\frac{u^4}{\C^2})$ and $\mathcal{O}(\frac{l^4}{\C^2})$, we obtain the following series solutions corresponding to $z_1(u)$ and $z_3(u)$. This gives
\begin{align}
&z_1(u)=\C^{\frac{5}{2}}\paren{\frac{-l^2}{2\C}+\frac{u^2}{2\C}+\frac{2}{5}\frac{l^4}{\C^2}-\frac{l^2u^2}{4\C^2}-\frac{3}{20}\frac{u^4}{\C^2}} \\
    &z_3(u)=(\C^{3/2}\left(\frac{u^2}{2\C}-\frac{l^2}{2\C}+\frac{1}{20}\frac{u^4}{\C^2}+\frac{l^4}{5\C^2}-\frac{1}{4}\frac{l^2u^2}{\C^2}\right)~.
\end{align}
Now with all these explicit forms of the RT surface for a universe with co-existing radiation and matter, we can further proceed to compute the area of the minimal RT surface. A perturbative calculation up to $\mathcal{O}(\m)$ and $\mathcal{O}(\T\chi)$ gives the following result for the area functional in eq.\eqref{RT_area_rad_ex}, which gives
\begin{align}
    A^{rad+ex}=\Omega_2\Bigg[&\frac{l^3}{2\z^2\sqrt{\C}}+\m\C^2\paren{\frac{1}{10}\frac{l^5}{\C^\frac{5}{2}}} \nonumber \\&+\T\chi \C\paren{\frac{1}{10}\frac{l^5}{\C^\frac{5}{2}}}\Bigg]~.
\end{align}
According to the RT formula, the holographic entanglement entropy is given by
\begin{align}\label{HEE rad+ex final}
    S_{HEE}^{(rad+ex)}=\frac{\Omega_2}{4\pi G_5}\Bigg[&\frac{l^3}{2\z^2\sqrt{\C}}+\m\C^2\paren{\frac{1}{10}\frac{l^5}{\C^\frac{5}{2}}} \nonumber \\&+\T\chi \C\paren{\frac{1}{10}\frac{l^5}{\C^\frac{5}{2}}}\Bigg]~.
\end{align}
Just like the previous two cases, here we will also compute the holographic entanglement entropy in the early and late time regimes of this universe with co-existing radiation and exotic matter. In order to obtain an early time behavior of the HEE, we will use the time-dependent brane position from eq.\eqref{early brane rad+ex}. Upon putting the brane position from eq.\eqref{early brane rad+ex} in the above expression for HEE, we finally obtain the following relation up to the leading order in $\tau$. This reads
\vspace{-0.02cm}
\begin{align}
    &A_{early}^{rad+ex}=\Omega_2\Bigg[\frac{l^3}{2z_i^3\sqrt{l^2+z_i^2}}\Bigg\{1+\sqrt{\m}\paren{z_i^2-\frac{\T{z_t}^2}{2}}\paren{\frac{3z_i^2+l^2}{z_i^2+l^2}}\; \tau\Bigg\}\nonumber\\&+\m\frac{l^5}{10\ci^{1/2}}\paren{1+\frac{1}{2}\frac{\sqrt{\m}z_i^2\paren{z_i^2-2\T z_t^2}}{\ci}\tau}+\T\chi \frac{l^5}{10\ci^{3/2}}\paren{1+\frac{3\;\sqrt{\m} z_i^2\paren{z_i^2-2\T z_t^2}}{2\ci}\tau}\Bigg]~.
\end{align}
Our result of HEE in the early time shows that the leading order behavior is of the $\mathcal{O}(\tau)$. Also this expression clearly shows the corrections in the HEE due to the presence of both radiation and exotic matter densities. As usual, the brane position and the HEE changes at the same rate at the early time in the leading order of the cosmological time $\tau$. We can also observe that in the early times of this universe, the result of HEE is dominated by the contribution from radiation. This is indeed consistent with the thermal history of the universe \cite{WMAP:2010qai,WMAP:2010sfg,Planck:2014loa,Planck:2018vyg}. Now we will proceed to find the late-time behavior of the HEE for a universe with coexisting radiation and exotic matter. In order to do so, we will put the late time, the brane position of eq.\eqref{late brane rad+ex} in eq.\eqref{HEE rad+ex final}, this gives 
\begin{align}
    &S^{rad+ex}_{late}=\frac{A_{AdS}}{4 G_{5}}+\frac{\m}{4 G_{5}} l^4 \Bigg[\frac{7}{10}-\frac{19}{20\T\chi l^2 \tau^2}\Bigg]+\frac{\T\chi}{4 G_{5}}l^2\Bigg[\frac{7}{10}-\frac{33}{20\T\chi l^2 \tau^2}\Bigg]
\end{align}
where 
\begin{equation}
\begin{aligned}
&A_{AdS}=\frac{l^{2}\left(3 z_i^{2}-4 z_t^{2}\right)\tau \chi}{8 \sqrt{m}\, z_i^{2} z_t^{2}}
+ \frac{1}{2} l^{2} \tau^{2} \chi
+ \frac{1}{2} l^{3}
\left(
\frac{
6\left(\frac{1}{2 \sqrt{m}\, z_i^{2}}-\frac{3}{8 \sqrt{m}\, z_t^{2}}\right)^{2}
+ \frac{2}{z_t^{2}\chi}
}{2l}
- \frac{1}{2 l^{3}\chi}
\right)\chi \\[6pt]
&+ \frac{
l^{2}\left(3 z_i^{2}-4 z_t^{2}\right)
\left(
48 m z_i^{4} z_t^{2}
+ 9 z_i^{4}\chi
- 24 z_i^{2} z_t^{2}\chi
+ 16 z_t^{4}\chi
\right)
}{
256 m^{3/2} z_i^{6} z_t^{6} \tau
} ~.
\end{aligned}
\end{equation}
The above expression clearly states that in the late time of a universe with coexisting radiation and exotic matter, HEE changes as $\tau^2$ in the leading order. As the brane position is proportional to $\tau$ in the late time, the leading order time dependence of HEE clearly suggests an area law behaviour. The subleading terms shows the corrections to the HEE due to the presence of coexisting matter sources in the universe.
\section{Holographic subregion complexity}\label{complexity section}
In this section, we will compute the holographic subregion complexity (HSC) for the FLRW universe with coexisting matter components. We shall also compute the early and late time behaviour of the HSC for different universes with coexisting matter components like radiation-dark matter and radiation-exotic matter. We would also like to mention that our calculations are done for a spherical subsystem on the brane. Before going into the details of the calculation, it is important to discuss about different holographic conjectures to compute the complexity of the boundary theory. \\
In the studies \cite{Susskind:2018pmk,Susskind:2014moa,Susskind:2014rva}, it is suggested that the complexity of the boundary states corresponds directly to the volume of the Einstein-Rosen bridge (ERB) connecting the boundaries of a two-sided eternal black hole. According to this proposal, the complexity of the boundary state is given by
\begin{equation}
    C_{V}(t_L, t_R)=\frac{V^{ERB}(t_L,t_R)}{8\pi RG_{d+1}}
\end{equation}
where $R$ is the AdS radius and $V^{ERB}(t_L,t_R)$ is the co-dimension one extremal volume of the Einstein-Rosen bridge. This Einstein-Rosen bridge is bounded by the two spatial slices at times $t_L$ and $t_R$ of two different CFTs that lie on the boundaries of an eternal black hole. The above equation is the well known as the ``Complexity=Volume'' or ``Complexity=Volume (1.0)'' conjecture in the literature.\\
There is another conjecture known as the ``Complexity=Volume (2.0)'' \cite{Stanford:2014jda,Alishahiha:2015rta}. This conjecture tells that the quantum complexity of the strongly interacting boundary CFT is dual to the volume underneath the minimal RT surface, which is extended in the bulk direction. Calculating the volume enclosed by the minimal RT surface and the boundary basically tells how difficult it is to construct the bulk degrees of freedom enclosed by the RT surface. According to this proposal, the quantum complexity of the boundary theory is given by 
\begin{equation}
    C_V =\frac{V_{\gamma}}{8\pi R G_{d+1}}
\end{equation}
where $V_{\gamma}$ is the volume enclosed by a minimal hypersurface and the boundary in the bulk. Some recent studies on the volume complexity in different bulk spacetime backgrounds including the FLRW, can be found in \cite{Ben-Ami:2016qex,An:2019opz,ChowdhuryRoy:2022dgo, Saha:2021kwq,Caginalp:2019fyt,Noumi:2025cup}.\\
Another famous conjecture tells that in order to compute the complexity, one needs to evaluate the bulk action on the Wheeler-DeWitt patch, which is enclosed by light sheets. This conjecture is referred to as the ``Complexity=Action'' conjecture \cite{Alishahiha:2018lfv,Brown:2015lvg,Brown:2015bva,Goto:2018iay}. According to this conjecture, the complexity of the boundary theory is given by 
\begin{equation}
    C_A =\frac{I_{WdW}}{\pi\hbar}~.
\end{equation}
In this paper, we will use the ``Complexity = Volume (2.0)'' conjecture to compute the quantum complexity of the spherical subsystem on the brane where the FLRW universe is situated. We will now proceed to compute the time-dependent HSC for the four-dimensional FLRW universe on the brane.
\subsection{Universe with no matter}
In this subsection, we will compute the time-dependent volume complexity of a universe with no matter field. This kind of universe is also referred to as an eternally inflating universe. In our RS-II braneworld model, to study a universe with no matter source, we will choose the pure AdS$_5$ spacetime as the bulk spacetime. \\
Now the main goal is to obtain the volume enclosed by the RT surface and the brane. Therefore, the volume underneath the RT surface is given by 
\begin{equation}
    V^{AdS}=\Omega_2 \int_{0}^{l} du u^2 \int_{z(u)}^{\z}\frac{dz}{z^4}=\Omega_2 \int_{0}^{l}du u^2 \Big(\frac{1}{3z(u)^3}-\frac{1}{\z^2}\Big)~.
\end{equation}
Now we will use the expression of $z(u)$ from eq.\eqref{RT surface profile AdS} in the above expression. This leads to the following result of the above-mentioned integral 
\begin{equation}
    V^{(AdS)}\approx\Omega_{2}\left[\frac{l^3}{9\C^{3/2}}+\frac{l^5}{10\C^{5/2}}+\frac{5l^7}{56\C^{7/2}}-\frac{l^3}{9\z^3}\right]~.
\end{equation}
It should be mentioned that while evaluating the above integral, we have neglected the terms higher than $\mathcal{O}(\frac{l^7}{\C^{7/2}})$.\\
Therefore, according to the ``Complexity=Volume'' conjecture, the volume complexity of the spherical subsystem of radius $l$ on the brane is given by
\begin{equation}\label{C AdS gen}
    C^{(AdS)}_{V}=\frac{\Omega_2}{8\pi G_5}\left[\frac{l^3}{9\C^{3/2}}+\frac{l^5}{10\C^{5/2}}+\frac{5l^7}{56\C^{7/2}}-\frac{l^3}{9\z^3}\right]~.
\end{equation}
With this expression of holographic volume complexity in hand, we will now proceed to compute its early and late time behaviour. As discussed earlier, the time dependence will enter through the brane position, which is a function of the cosmological time.\\
In the early time, we can obtain a time-dependent expression for the holographic volume complexity by substituting the brane position from eq.\eqref{early brane pos AdS} in the above equation. This gives a time-dependent expression for the holographic volume complexity up to $\mathcal{O}(\tau)$, which reads 
\begin{align}
    C^{(AdS)}_{V}&=\frac{\Omega_{2}}{8\pi G_{5}}\Bigg[\frac{l^3 r_{0}^3}{9(l^2 r_{0}^2 +1)^{3/2}}\paren{1+\frac{3 H\tau}{l^2 r_{0}^2 +1}}+\frac{l^5 r_{0}^5}{10(l^2 r_{0}^2 +1)^{5/2}}\paren{1+\frac{5 H\tau}{l^2 r_{0}^2 +1}}\nonumber\\
    &\quad\Bigg.+\frac{5l^7 r_{0}^7}{56(l^2 r_{0}^2 +1)^{3/2}}\paren{1+\frac{3 H\tau}{l^2 r_{0}^2 +1}}-\frac{l^3 r_{0}^3}{9}(1+3H\tau)\Bigg]~.
\end{align}
From the above expression of $C^{(AdS)}_{V}$, it is clear that the holographic volume complexity of an eternally inflating universe changes as $\tau$ in the early time limit. Thus, both the brane position and complexity change at the same rate in the early time of an eternally inflating universe.\\
Again, in the late time (that is, $\tau\to\infty$), the brane position obeys the mathematical expression given in eq.\eqref{late brane pos AdS}. Upon substituting this equation (eq.\eqref{late brane pos AdS}) in eq.\eqref{C AdS gen}, we finally arrive at the following late-time expression for the holographic volume complexity
\begin{align}
    C^{(AdS)}_{V}&=\frac{\Omega_{2}}{8\pi G_{5}}\Bigg[\frac{1}{9}\paren{1-\frac{3e^{-2H\tau}}{2l^2 r_{0}^2}}+\frac{1}{10}\paren{1-\frac{5e^{-2H\tau}}{2l^2 r_{0}^2}}+\frac{5}{56}\paren{1-\frac{7e^{-2H\tau}}{2l^2 r_{0}^2}}-\frac{l^3 r_{0}^3}{9}e^{3H\tau}\Bigg]~.
\end{align}
The above equation clearly states that in the late-time regime, the leading-order time dependence of volume complexity is $e^{3H\tau}$. As we already know that the actual length scale $L\propto e^{H\tau}$, the complexity follows a volume law, which is expected in the late-time regime.
\subsection{Universe with coexisting radiation and matter}
In this subsection, we will compute the time-dependent volume complexity for a universe with coexisting radiation and matter. We shall also compute the early and late time behaviour of the volume complexity using the time-dependent brane position for both early and late time scenarios.\\
The volume under the static RT surface is given by
\begin{equation}
    V^{(rad+mat)}=\Omega_{2}\int_{0}^{l}du u^2 \int_{z(u)}^{\z}\frac{dz}{z^4 \sqrt{f(z(u))}}
\end{equation}
where $f(z(u))=1-\T\rc z(u)^3 -\m z(u)^4$. \\
Using this lapse function in the above expression for volume and evaluating the integral up to $\mathcal{O}(\m)$ and $\mathcal{O}(\T\rc)$, we get 
\begin{align}
    V^{(rad+matt)}& \approx\Omega_{2}\left[\left(\frac{l^{3}}{9\C^{3/2}}+\frac{l^5}{10 \C^{5/2}}+\frac{5 l^7}{56 \C^{7/2}}+\frac{g l^3}{3}\right)\right.\nonumber \\&\left. +\m \C^{2}\left(-\frac{l^3}{6 \C^{3/2}}+\frac{7 l^5}{60 \C^{5/2}}+\frac{7 l^7}{240 \C^{7/2}}\right)\right.\nonumber \\&\left. 
    +\T{\rc} \C^{3/2} \left(\frac{7}{60}\frac{l^5}{\C^{5/2}}+\frac{9}{280}\frac{l^7}{\C^{7/2}}-\frac{1}{6}\frac{l^3}{\C^{3/2}}ln\left(\sqrt{\C}\right)\right)\right]~.
\end{align}
where $g\equiv -\frac{1}{3\z^3}+\frac{\m}{2}\z+\frac{\T\rc}{2}\ln{\z}$. \\
Therefore, from the ``Complexity=Volume'' conjecture, the complexity of the spherical subregion of radius $l$ of the FLRW universe is given by
\begin{align}\label{complexity rad+mat gen}
    C_{V}^{(rad+matt)}& =\frac{\Omega_2}{8\pi G_5}\left[\left(\frac{l^{3}}{9\C^{3/2}}+\frac{l^5}{10 \C^{5/2}}+\frac{5 l^7}{56 \C^{7/2}}+\frac{g l^3}{3}\right)\right.\nonumber \\&\left. +\m \C^{2}\left(-\frac{l^3}{6 \C^{3/2}}+\frac{7 l^5}{60 \C^{5/2}}+\frac{7 l^7}{240 \C^{7/2}}\right)\right.\nonumber \\&\left. 
    +\T{\rc} \C^{3/2} \left(\frac{7}{60}\frac{l^5}{\C^{5/2}}+\frac{9}{280}\frac{l^7}{\C^{7/2}}-\frac{1}{6}\frac{l^3}{\C^{3/2}}ln\left(\sqrt{\C}\right)\right)\right]~.
\end{align}
With the expression of volume complexity in hand, we will now proceed to compute the early and late time behaviour of complexity for a universe with co-existing radiation and dark matter. In order to find the early time behaviour of the volume complexity, we will use the early time brane position of eq.\eqref{early_brane_rad+matt} in the above expression for the subregion volume. This gives the following time-dependent expression
\begin{align}
  C_{V_{early}}^{(rad+matt)}=&\frac{\Omega_2}{8\pi G_5}\Big[a+b\tau-\rc \frac{l^3}{6}  \ln\Bigg\{\ci\paren{1-\frac{\sqrt{\m}\;z_i^2}{l^2+z_i^2}\paren{z_i^2+\frac{z_iz_t}{2}}\tau}\Bigg\}\nonumber \\&+\frac{\T\rc}{2}\ln{z_i\paren{1-\sqrt{\m}\;\paren{z_i^2+\frac{z_iz_t}{2}}\tau}}\Big]
\end{align}
where
\begin{align}
    a&=\frac{l^3}{9\ci^{3/2}}+\frac{l^5}{10\ci^{5/2}}+\frac{5l^7}{56\ci^{7/2}}-\frac{l^3}{9z_i^3}\nonumber\\&+\frac{\m\ci^{2}}{2}\bigg(\frac{l^3}{6\ci^{3/2}}+\frac{7l^5}{60\ci^{5/2}}+\frac{7l^7}{240\ci^{7/2}}\bigg)+\frac{\m l^3z_i}{6}\nonumber\\&+\rc \ci^{3/2}\bigg(\frac{7l^5}{60\ci^{5/2}}+\frac{9l^7}{280\ci^{7/2}}\bigg)
\end{align}
\begin{align}
    b&=\frac{\sqrt{\m}z_i^2}{l^2+z_i^2}\bigg(z_i^2+\frac{z_i z_t}{2}\bigg)\Bigg(\bigg(\frac{l^{3}}{3\ci^{3/2}}+\frac{l^{5}}{2\ci^{5/2}}+\frac{5l^{7}}{8\ci^{7/2}}\bigg)-\frac{l^3\ci}{3z_i^5}-\frac{\m\ci}{2z_i}\nonumber\\&+\m\ci^{2}\bigg(\frac{-l^3}{6\ci^{3/2}}+\frac{7l^5}{60\ci^{5/2}}+\frac{7l^7}{6\ci^{7/2}}\bigg)\nonumber\\&+\rc\ci^{3/2}\bigg(\frac{7l^5}{30\ci^{5/2}}+\frac{9l^7}{70\ci^{7/2}}\bigg)\Bigg)~.
\end{align}
From the above expression, it is clear that for the early time of a universe with coexisting radiation and matter, the volume complexity changes as $\tau$ in the leading order. This is a clear indication of radiation domination in the result of volume complexity in the early times of the universe, with coexisting radiation and matter. Our result clearly shows correction terms containing both radiation and matter densities, which is quite expected for universes with coexisting matter sources. \\
Now we will compute the time-dependent volume complexity for the late time of a universe with coexisting radiation and matter. We will now use the expression of the time-dependent brane position for the late time regime of this universe from eq.\eqref{late_brane_rad+matt}. After substituting eq.\eqref{late_brane_rad+matt} in eq.\eqref{complexity rad+mat gen}, we finally obtain the following relation for the volume complexity
\begin{align}
& C_{V_{late}}^{rad+matt}=\frac{\Omega_2}{8\pi G_5}\Bigg[\frac{757}{2520}-\frac{l^{4} \T\rc}{48 z_t}
+\frac{25 l^{3} r}{168}
- \frac{1}{6} l^{3} r \log(l)- \frac{l^{3} \tau^{2}}{9 \alpha^{3}}+\frac{l^{3} \beta \tau^{4/3}}{3 \alpha^{3}}- \frac{l^{3} \gamma \tau}{3 \alpha^{3}}\nonumber\\&
+ \frac{1}{6} l^{3} r \log\!\left( \frac{\alpha}{\tau^{2/3}} \right)- \frac{2 l^{3} \beta^{2} \tau^{2/3}}{3 \alpha^{3}}+ \frac{4 l^{3} \beta \gamma \tau^{1/3}}{3 \alpha^{3}}
+ \frac{10 l^{3} \beta^{3}}{9 \alpha^{3}}
- \frac{2 l^{3} \gamma^{2}}{3 \alpha^{3}}
- \frac{10 l^{3} \beta^{2} \gamma}{3 \alpha^{3} \tau^{1/3}}
\Bigg]
\end{align}
where $\alpha=\paren{\frac{2}{3}}^{2/3}\T\rc^{-1/3}$, $\beta=\paren{\frac{2}{3}}^{2/3}\T\rc^{-1/3}\frac{1}{z_t}$ and $\gamma=\frac{2\sqrt{z_t}}{3\sqrt{\T\rc}}\paren{\frac{2}{3z_t^2}-\frac{1}{2z_i^2}+\frac{z_t}{6z_i^3}}$.\\
The above equation clearly shows that in the late time of a universe with coexisting radiation and matter, the holographic volume complexity changes as $\tau^2$ in the leading order and changes as $\tau^{4/3}$ in the sub-leading order. As we already know that the physical length scale in the late time of a universe with coexisting radiation and matter changes as $\tau^{2/3}$ in the leading order, it is obvious from our result that complexity obeys a volume law in the leading order. The other sub-leading terms in the above expression contain both radiation and dark matter densities as expected. This result is also consistent with the thermal history of the universe \cite{WMAP:2010qai,WMAP:2010sfg,Planck:2014loa,Planck:2018vyg}.
\subsection{Universe with coexisting matter and radiation in intermediate regime}
\noindent 
In this subsection, we aim to evaluate the time-dependent holographic volume complexity of a universe containing both radiation and matter in an intermediate regime. As previously mentioned in subsection \eqref{sub sec 4.3}, the intermediate regime corresponds to the region where the ratio \(\tilde{m}/r^4\) is comparable to \(\rho_p / r^{3}\). In order to compute the time-dependent HSC in this scenario, we will use the previously obtained result of HSC from eq.\eqref{complexity rad+mat gen}. After substituting the time-dependent radial position of the brane ($\z=z_t(1-\epsilon)$) in the intermediate regime from eq.\eqref{z bar inter} into eq.\eqref{complexity rad+mat gen}, and retaining terms up to $\mathcal{O}(\epsilon)$, we obtain the following expression for the HSC up to $\mathcal{O}(\epsilon)$ 
\begin{equation}\label{HSC inter}
    C^{rad+mat}_{int}\sim \frac{\Omega_2}{24\pi G_5}\frac{l^3}{z_t^3}(1+3\epsilon)~.
\end{equation}
We already know that the actual length scale ($L$) measured by an observer sitting outside the co-moving frame changes as $L\sim\frac{l}{z_t}(1+\epsilon)$. Therefore, the above expression of $C^{rad+mat}_{int}$ scales as $L^3$, which is the volume law.\\
Now we will proceed further to evaluate the explicit time dependence of the HSC in the intermediate regime where both radiation and matter play an important role. After substituting the form of $\epsilon$ from eq.\eqref{epsilon1} into eq.\eqref{HSC inter}, we get
\begin{equation}\label{HSC inter time}
    C^{rad+mat}_{int}\sim \frac{\Omega_2}{24\pi G_5}\frac{l^3}{z_t^3}\Bigg[1+3\Bigg(\frac{1}{3}\Big(\frac{2}{\m}\Big)^{\frac{3}{2}}\Big(\frac{3}{2}\T\rc^2\tau - \T C\Big)+\frac{4}{3}\Bigg)\Bigg]~.
\end{equation}
\subsection{Universe with coexisting radiation and exotic matter}
Now we will compute the holographic volume complexity and its early and late time behaviour for a universe with co-existing radiation and some kind of exotic matter. Just like the previous subsection, we have to first compute the volume under the RT surface. We know that the volume enclosed by the RT surface and the brane is given by 
\begin{equation}
    V^{(rad+ex)}=\Omega_{2}\int_{0}^{l}du u^2 \int_{z(u)}^{\z}\frac{dz}{z^4 \sqrt{f(z(u))}}
\end{equation}
where $f(z(u))=1-\T\chi z(u)^2 -\m z(u)^4$. \\
The above integral can be easily evaluated in a perturbative way. By expanding all the terms up to an $\mathcal{O}(\T\chi)$ and $\mathcal{O}(\m)$ and keeping terms up to $\mathcal{O}(\frac{l^7}{\C^{7/2}})$, we finally obtain
\begin{align}
    V^{(rad+ex)}& =\Omega_{2}\left[\left(\frac{l^{3}}{9\C^{3/2}}+\frac{l^5}{10 \C^{5/2}}+\frac{5 l^7}{56 \C^{7/2}}+\frac{g l^3}{3}\right)\right.\nonumber \\&\left. +\m \C^{2}\left(-\frac{l^3}{6 \C^{3/2}}+\frac{7 l^5}{60 \C^{5/2}}+\frac{l^7}{240 \C^{7/2}}\right)\right.\nonumber \\&\left. 
    +\T\chi\;\C\paren{\frac{l^3}{6\C^{3/2}}+\frac{7}{60}\frac{l^5}{\C^{5/2}}+\frac{37}{1680}\frac{l^7}{\C^{7/2}}}\right]~.
\end{align}
Therefore, according to the ``Complexity=Volume'' conjecture, the holographic volume complexity of a circular subsystem on the brane for a universe with co-existing radiation and exotic matter is given by
\begin{align}\label{comp rad+ex gen}
    C_{V}^{(rad+ex)}& =\frac{\Omega_{2}}{8\pi G_5}\left[\left(\frac{l^{3}}{9\C^{3/2}}+\frac{l^5}{10 \C^{5/2}}+\frac{5 l^7}{56 \C^{7/2}}+\frac{g l^3}{3}\right)\right.\nonumber \\&\left. +\m \C^{2}\left(-\frac{l^3}{6 \C^{3/2}}+\frac{7 l^5}{60 \C^{5/2}}+\frac{l^7}{240 \C^{7/2}}\right)\right.\nonumber \\&\left. 
    +\T\chi\;\C\paren{\frac{l^3}{6\C^{3/2}}+\frac{7}{60}\frac{l^5}{\C^{5/2}}+\frac{37}{1680}\frac{l^7}{\C^{7/2}}}\right]~.
\end{align}
Now the only task left is to evaluate the early and late time behaviour of the volume complexity mentioned in the above equation. At first, we will use the early time brane position of eq.\eqref{early brane rad+ex}, in the above expression for volume complexity. After putting the time-dependent brane position from eq.\eqref{early brane rad+ex} and expanding terms up to the leading order in $\tau$, we get
\begin{align}
 C_{V_{early}}^{(rad+ex)}=\frac{\Omega_2}{8\pi G_5}\Bigg[c+d\tau+\frac{l^3\rc}{6}lnz_i\bigg(1-\sqrt{\m}\big(z_i^2-\frac{\Tilde z^2_t}{2}\big)\tau\bigg)\Bigg]
\end{align}
Where,

\begin{align}
c&=\frac{l^3}{9\ci^{3/2}}+\frac{l^5}{10\ci^{5/2}}+\frac{5l^7}{56\ci^{7/2}}-\frac{l^3}{9z_i^{3}}+\frac{l^3\m z_i}{6}\nonumber\\&+\m\ci^{2}\Bigg(-\frac{l^3}{6\ci^{3/2}}+\frac{7l^5}{60\ci^{5/2}}+\frac{l^7}{240\ci^{7/2}}\Bigg)\nonumber\\&+\T\chi\ci\Bigg(\frac{l^3}{6\ci^{3/2}}+\frac{7l^5}{60\ci^{5/2}}+\frac{37l^7}{1680\ci^{7/2}}\Bigg)
\end{align}
\begin{align}
d&=\frac{\sqrt{\m}z_i^2}{l^2+z_i^2}\bigg(z_i^2+\frac{z_i z_t}{2}\bigg)\Bigg[\frac{l^3}{3\ci^{3/2}}+\frac{l^5}{2\ci^{5/2}}+\frac{5l^7}{8\ci^{7/2}}-\frac{l^3}{3}\bigg(\frac{\ci}{z_i^5}+\frac{\m\ci}{2z_i}\bigg)\nonumber\\&+\m\ci^{2}\bigg(-\frac{l^3}{2\ci^{3/2}}+\frac{7l^5}{12\ci^{5/2}}+\frac{7l^{7}}{240\ci^{7/2}}\bigg)\nonumber\\&-4\m\ci^{2}\bigg(-\frac{l^3}{6\ci^{3/2}}+\frac{7l^5}{60\ci^{5/2}}+\frac{l^7}{240\ci^{7/2}}\bigg)\nonumber\\&-2\T\chi\ci\bigg(\frac{l^3}{6\ci^{3/2}}+\frac{7l^5}{60\ci^{5/2}}+\frac{37l^7}{1680\ci^{7/2}}\bigg)\nonumber\\&+\T\chi\ci\bigg(\frac{l^{3}}{2\ci^{3/2}}+\frac{7l^5}{12\ci^{5/2}}+\frac{37 l^7}{240\ci^{7/2}}\bigg)\Bigg]~.
\end{align}
The above equation clearly shows that holographic volume complexity changes as $\tau$ in the leading order. Similar to our earlier observations, here we can also state that volume complexity for a universe with co-existing radiation and exotic matter changes at the same rate as the brane position in the early time regime. \\
For the late time, that is in the $\tau\to\infty$, the brane position changes as eq.\eqref{late brane rad+ex} for a universe with co-existing radiation and exotic matter. After substituting the mathematical form of the brane position from eq.\eqref{late brane rad+ex} in eq.\eqref{comp rad+ex gen}, we get the following late time behaviour for the volume complexity
\begin{align}
&C_{V_{late}}^{(rad+ex)}=\frac{\Omega_2}{8\pi G_5}\Bigg[\frac{19}{90}
+ \frac{5}{56\, l^{2}}- \frac{\T\chi}{\T z_t^2}\frac{11 \; l^4}{240}+\T\chi \frac{171\;l^2}{560}- \frac{1}{6} l^{3}\,\beta\,\T\chi^{3/2}- \frac{10}{9} l^{3}\,\beta^{3}\,\T\chi^{3/2}
\nonumber \\&- \frac{1}{9} l^{3}\,\tau^{3}\,\T\chi^{3/2}
- \frac{1}{3} l^{3}\,\beta\,\tau^{2}\,\T\chi^{3/2}
- \frac{1}{6} l^{3}\,\tau\,\T\chi^{3/2}
- \frac{2}{3} l^{3}\,\beta^{2}\,\tau\,\T\chi^{3/2}
- \frac{1}{3} l^{3}\,\gamma\,\tau\,\T\chi^{3/2}
+ \frac{l^{3}\,\sqrt{\T\chi}}{6\, \T z_t^{2}\,\tau}
\nonumber \\ &
- \frac{l^{3}\,\beta^{2}\,\T\chi^{3/2}}{6\,\tau}
- \frac{5\, l^{3}\,\beta^{4}\,\T\chi^{3/2}}{3\,\tau}
- \frac{l^{3}\,\gamma\,\T\chi^{3/2}}{6\,\tau}
- \frac{10\, l^{3}\,\beta^{2}\,\gamma\,\T\chi^{3/2}}{3\,\tau}
- \frac{2\, l^{3}\,\gamma^{2}\,\T\chi^{3/2}}{3\,\tau}
\Bigg]
\end{align}
where $\beta=\frac{\T z_t}{\sqrt{\T\chi}}\paren{\frac{3}{8\T z_t^2}-\frac{1}{2z_i^2}}\frac{1}{\tau}$ and $\gamma=\frac{1}{2\chi\T z_t^2}$.\\
By looking at the above equation, one can tell that in the late time holographic volume complexity changes as $\tau^3$ and the sub-leading term of cosmological time is of the order $\tau^2$. As we have both radiation and exotic matter in our universe, it is quite obvious to expect the complexity will depend on both radiation and exotic matter density $\m$ and $\T\chi$, respectively. Our expression for the volume complexity clearly shows the presence of radiation and exotic matter density as expected. Another important observation is that the physical length scale changes as $L\propto\tau$ in the leading order. Therefore, holographic complexity changes as $\tau^3$ and follows a volume law in the leading order. We can also say that the late-time behavior of volume complexity has a dominant contribution from the exotic matter component in the late time, which is consistent with the thermal history of a universe \cite{WMAP:2010qai,WMAP:2010sfg,Planck:2014loa,Planck:2018vyg} with co-existing radiation and exotic matter.

\section{Conclusion}
We conclude our article with a summary of the key findings and an overview of the work presented in this article. In Section 2,  we have given a detailed derivation of the Israel junction condition, which gives a differential equation of the brane position with time for a general black brane metric. In section 3, we have added a string action to add different kinds of matter in the system and have shown that it is possible to introduce a lapse function corresponding to not only one component but also a two-component universe. This kind of lapse function can be used to describe a more general, physical cosmological scenario, as from the thermal history, it is well established that the universe we live in has more than one component domination in the different eras since the start of the universe. Among all theoretical models, radiation and matter have been observationally proven. Using the Israel junction condition and integrating, we have determined the brane position as a function of time for a two-component universe. Finding this time dependence is very important because in the later section, this determines the time dependence of the holographic entanglement entropy and complexity. The next sections are dedicated to the calculation of the holographic entanglement entropy and holographic complexity using a perturbative procedure developed in \cite{Paul:2025gpk}.\\ 
The motivation for the study of the entanglement measures of the universe using AdS/CFT duality comes from the fact that just after the Big Bang, a strongly coupled phase came, and a phase transition like QCD, of confinement and de-confinement, happened. While using the HRT formalism is natural for the non-static FLRW background, we have used the RT formalism because the HRT prescription is very hard to implement, as it deals with non-linear and coupled differential equations. Instead, the RT formalism is easier, and it also matches the results of the HRT formalism in the leading order. To study a universe expanding as a power law of time, the brane world (RS-II) prescription has been used, where the four-dimensional universe has been taken to be situated on a brane, and the expansion is realised by the radial motion of the brane in the bulk direction. Different matters on the braneworld are the result of back reaction caused by different $p$-brane gas geometries in the bulk spacetime. On the other hand, while calculating the HSC, we have taken the ``Complexity=Volume'' conjecture. In both the HEE and HSC, we first derive these quantities in terms of the brane position, and the time dependence enters through the time dependency of the brane position. We have calculated the time-dependent HEE and HSC for an eternally inflating (same as exponential late time expansion) universe, a universe in the presence of both radiation and matter, and a universe with radiation and exotic matter. We have found the HEE and HSC for all of these for early and late times. \\
It has been demonstrated that, in the absence of any matter sources, the holographic entanglement entropy (HEE) scales linearly with the cosmic time ($\tau$) during the early universe, and exponentially as $e^{2H\tau}$ at late times. This behavior aligns perturbatively with the findings of \cite{Park:2020jio}. For the most realistic model that is a universe in the presence of both radiation and matter, after analysing, we found that the early time dependence of the HEE scales linearly in leading order and also some non-zero contribution from the dark matter sector is present there. Through a similar calculation, we found in the late time that HEE scales with time as $\tau^{4/3}$ and this is not solely due to the dark matter sector but radiation contribution through the term $z_t$. So in the early time there was radiation domination, and in the late time there is matter domination, and hence in the early time radiation part contributes in the leading order, but matter contributes in the sub-leading order, and in the late time it is the opposite. \\
In the very next section, we have shown that for pure AdS, the HSC scales linearly with the cosmic time in the leading order in early times and in the late time it scales as $e^{3H\tau}$, matching the expectation that as the length scale is growing as $e^{H\tau}$, the HSC is growing as length to the power of three. For the case of the universe filled by dark matter and radiation, the HSC in the early time grows linearly with time having the dominant contribution from radiation and sub-dominant contribution from the matter sector; in the late time the HSC scales as $\tau^2 $ in the leading order and $\tau^{4/3}$ in the sub-leading order, where the leading order contribution is clearly coming from the matter sector and the sub-leading order contribution is due to matter and radiation both. We have also investigated the intermediate regime of coexistence of radiation and matter where $\frac{\m}{r^4}\sim\frac{\rho_1}{r^3}$, and obtained analytical expressions for the HEE and HSC. We found that these quantities obey the area and volume law, respectively, in the intermediate regime. In a universe filled with radiation and exotic matter, our early-time calculations of the HSC reveal that the leading-order behavior scales linearly with \(\tau\), with a contribution arising from the exotic matter in the subleading order. At late times, the dominant correction from the exotic matter scales as \(\tau^3\), while the sub-leading correction from radiation scales as \(\tau^2\).
We are concluding this article by noting that this study shows that it is possible to proceed with a more general and physical universe to calculate other quantum information-theoretic measures like entanglement negativity (EN), entanglement wedge cross section (EWCS), mutual information (MI), etc. We are leaving these computations in the braneworld scenario for future work.
\section*{Acknowledgment} 
RM would like to thank SNBNCBS for the Junior Research Fellowship. 
GG extends his heartfelt gratitude to CSIR, Govt. of India, for funding this research. SP would like to thank SNBNCBS for the Fellowship.  The authors would also like to thank the anonymous referee for useful comments and suggestions.\\
\bibliographystyle{hephys}
\bibliography{refptep}

\end{document}